\documentstyle[here,epsfig]{mn2e}
\def\simlt{\mathrel{\rlap{\lower 3pt\hbox{$\sim$}}\raise 2.0pt\hbox{$<$}}}
\def\simgt{\mathrel{\rlap{\lower 3pt\hbox{$\sim$}} \raise 2.0pt\hbox{$>$}}}

\def\gtsima{$\; \buildrel > \over \sim \;$}
\def\ltsima{$\; \buildrel < \over \sim \;$}
\def\gtrsim{\lower.5ex\hbox{\gtsima}}
\def\lesssim{\lower.5ex\hbox{\ltsima}}
\begin{document}

\newcommand{\q}{\begin{equation}}
\newcommand{\qa}{\begin{eqnarray}}
\newcommand{\qs}{\begin{eqnarray*}}
\newcommand{\nq}{\end{equation}}
\newcommand{\nqa}{\end{eqnarray}}
\newcommand{\nqs}{\end{eqnarray*}}
\newcommand{\ud}{\mathrm{d}}

\title[Background radiation from sterile neutrino decay and reionization] 
{Background radiation from sterile neutrino decay and reionization}
%{Sterile neutrino decays: effects on the background radiation and on the reionization} 
\author[M. Mapelli \& A. Ferrara] 
{M. Mapelli \&  A. Ferrara \\
SISSA, Via Beirut 4, 34100, Trieste, Italy\\}

\maketitle \vspace {7cm }

\begin{abstract}
Sterile neutrinos are one of the most promising Warm Dark Matter candidates. 
By considering their radiative- and pion-decay channels,  
we derive the allowed contribution of sterile neutrinos to the X-ray, optical 
and near-infrared cosmic backgrounds. The X-ray background puts a strong 
constraint on the mass of radiatively decaying neutrinos ($m_{\nu{}s}\lesssim{}14$ keV), 
whereas the allowed mass range for pion-decay neutrinos 
(for a  particle lifetime $> 4\times{}10^{17}$ s)
is $150 \le m_{\nu{}s}/{\rm MeV} \le 500$. 
Taking into account these constraints, we find that sterile neutrinos do not significantly contribute to the 
optical and near-infrared background. We further
consider the impact of sterile neutrinos on reionization.  
We find that the Thomson optical depth due to sterile neutrinos is $\tau_e =(0.4-3)\times{}10^{-2}$ in the
case of radiative decays, and it is $\approx{}10^{-3}$ for the
pion-decay channel.
We conclude that these particles must have played only a minor role in cosmic reionization history.

%Whereas radiatively decaying neutrinos produce a Thomson optical depth $0.11 \le \tau_e \le 0.13$, we find that $\tau_e \le 0.11$ for pion-decaying neutrinos in all cases. 
\end{abstract}

\section{Introduction}
%Part I: literature\\
Different dark matter candidates can be distinguished on the basis of their velocity dispersion, corresponding to a free-streaming length,
%approximately the speed of the HDM particles times the age of the universe
%\lambda{}_{FS}\sim{}5\times{}10^{-3}(\Omega{}_{DM}h^2)^{1/3}\left(\frac{m_{DM}}{1\textrm{GeV}}\right)^{-4/3}\textrm{pc}
 below which dark matter fluctuations are suppressed. In Cold Dark Matter (CDM) models particles have negligible free-streaming length with respect to scales of cosmological interest. On the contrary, in Warm Dark Matter (WDM) scenarios the velocity dispersion of the particles is sufficient to smear out  the fluctuations up to galactic scales, depending on the mass of the particles (Padmanabhan 1995). For this reason WDM models can alleviate the so called substructure crisis, which represents one of the most serious problems of CDM theories (Bode, Ostriker \& Turok 2001; Ostriker \& Steinhardt 2003).\\
%Sterile neutrinos are one of the most promising dark matter candidates in Warm Dark Matter (WDM) models (Sommer-Larsen \& Dolgov 2001; Bode, Ostriker \& Turok 2001).
 Sterile neutrinos are considered the most promising WDM candidates (Colombi, Dodelson \& Widrow 1996; Sommer-Larsen \& Dolgov 2001). They can exist only if neutrinos have non-zero mass and mixing angles, as predicted by the standard oscillation theory (Dolgov \& Hansen 2002; see Dolgov 2002 for a complete review of sterile neutrino properties). They are massive; so they can decay into lighter particles, following a large number of different possible channels (Dolgov 2002). Among all the possible decay processes, those which lead to the emission of photons are particularly interesting, because we can put constraints on the existence, the mass, the lifetime, and other properties of sterile neutrinos, depending on the detection of such radiation (De R\'ujula \& Glashow 1980; Stecker 1980; Drees \& Wright 2000; Abazajian, Fuller \& Patel 2001a; Abazajian, Fuller \& Tucker 2001b; Dolgov 2002 and reference therein). In principle, radiation due to neutrino decays should be seen both as line emission in the local universe (at energy $E_\gamma{}=(m_{\nu{}s}^2-m_{\nu{}a}^2)/2m_{\nu{}s}$, where $m_{\nu s}$ and $m_{\nu{}a}$ are respectively the mass of the sterile and of the light neutrino), and as background radiation, given by the integrated contribution of decays occurring at different redshifts. Until now, none of the experiments built to detect emission lines in the local universe has provided any evidence of the existence of sterile neutrinos (Henry \& Feldman 1981; Shipman \& Cowsik 1981;  Holdberg \& Barger 1985; Bowyer et al. 1999). The 
absence of detection can be used to fix upper limits on the mass and lifetime of sterile neutrinos.
In particular, measurements of X-ray emission from galaxy clusters (Abazajian, Fuller \& Tucker 2001b) provide a strong upper limit of the rest mass of radiatively decaying sterile neutrinos ($m_{\nu{}s}\lesssim{}5$ keV in the zero lepton production mode). This limit is particularly constraining, given that studies (based on Ly$\alpha{}$ forest data) of power spectrum fluctuations due to free-streaming effects provide a lower limit for the mass of sterile neutrinos $m_{\nu{}s}\gtrsim{}2$ keV (Viel et al. 2005; see also Narayanan et al. 2000
% for the previous estimate $m_{\nu{}s}\gtrsim{}0.75$ keV in the case of thermal neutrinos
\footnote{The lower limit of 0.75 keV for thermal neutrino mass derived by Narayanan et al. (2000) can be converted into a lower limit of $\sim{}2$ keV for sterile neutrino mass, as shown by Hansen et al.~(2002).}). 
%On the other hand, also 
Studies of cosmological contributions of sterile neutrinos to the background radiation in different bands (Stecker 1980; Kimble, Bowyer \& Jakobsen 1981; Ressel \& Turner 1990; Davidsen et al. 1991; Abazajian, Fuller \& Tucker 2001b) can only provide upper limits of their mass and lifetime. In addition, upper limits derived from background radiation can be biased by dark matter clustering (Abazajian, Fuller \& Tucker 2001b).

It has also been proposed that radiatively decaying sterile neutrinos can be responsible of the reionization (Sciama 1982; Rephaeli \& Szalay 1981; Sciama 1990; Sciama 1993).
This idea does not encounter particular success today, because there are no observational evidences supporting it.
% This idea has not been completely rejected; but there are no observational evidences supporting it. 
The main failure of this model is that the ionizing sources proposed by Sciama, sterile neutrinos of mass $m_{\nu{}s}=27.4\pm{}0.2$ eV, must be excluded on the basis of the lower limit derived by the power spectrum analysis (Narayanan et al. 2000; Viel et al. 2005).
% In addition, although this model has not been falsified yet, there are no observational evidences supporting it.
On the other hand, even if it seems unlikely that sterile neutrinos are the only responsible of the reionization, we cannot exclude that they partially contribute to it, together with other sources.
Furthermore, the high 
%Thomson 
electron scattering optical depth  ($\tau{}_e=0.17\pm{}0.04$) resulting by WMAP data (Spergel et al 2003; Kogut et al. 2003), combined with the indication that the universe was not fully ionized until redshift 6-7, as one can derive from the HI absorption present in the spectra of highest redshift quasars in the Sloan Digital Sky Survey (Becker et al. 2001; Fan et al. 2003), suggests a long (and, maybe, complex) reionization history, difficult to achieve considering only ''traditional sources'' (as Population II stars and quasars). One of the proposed scenarios (Cen 2003; Furlanetto \& Loeb 2004) predicts a first high redshift phase of partial ionization, triggered by quite exotic objects, followed by a most recent phase of complete and definitive re-ionization, ended at redshift about 6 and due to normal Population II stars and quasars.
Among the various candidates proposed as sources of the first partial ionization, like massive metal-free stars (Haiman \& Holder 2003; Choudhury \& Ferrara 2004) and micro-quasars (Madau et al. 2004), one could reasonably include also radiatively decaying neutrinos.
%WMAP results (Spergel et al 2003; Kogut et al. 2003), giving a considerable high Thomson optical depth ($tau=0.17$), combined with the analysis of the spectra of high redshift blazars (...), that indicate a neutral hydrogen fraction of the order of 1\% until redshift 6, seem to require a complex reionization history, difficult to achieve considering only traditional sources (like Population II stars and quasars).
 \\
%An alternative approach to detect also sterile neutrinos that do not lead to photon emission has recently been proposed by Haiman and Hansen (2003).
%As an alternative, Hansen \& Haiman (2004) have recently suggested a way to detect also sterile neutrinos that do not directly decay into photons. They noticed that
As an alternative, Hansen \& Haiman (2004, hereafter HH) have recently suggested that also sterile neutrinos which do not directly decay into photons can be a source of this partial ionization. They noticed that very massive sterile neutrinos ($140\leq{}m_{\nu{}s}\leq{}500$ MeV) mainly decay into pions and leptons (electrons or positrons). These electrons are sufficiently energetic to Compton-scatter CMB photons, which can then ionize hydrogen atoms directly or through secondary electrons.

%\section{METHOD}
In this paper we want to check the possible contribution of sterile neutrinos to the background radiation, in the light of the most recent measurements of the extragalactic background light (EBL), with particular care for the optical-infrared (Madau \& Pozzetti 2000; Bernstein et al. 2002; Matsumoto et al. 2000; Wright 2000) and for the X-ray band (Gruber 1992; Hornschmeier et al. 2001; Tozzi et al. 2001; Moretti et al. 2003 [M03]; Dijkstra et al. 2004 [D04]; Bauer et al. 2004 [B04]). We, further, discuss the role of radiatively and pion-decaying sterile neutrinos in the cosmic reionization, calculating the corresponding Thomson optical depth. 
%%%%%In particular, in Section 2 we present our calculations of the flux due to sterile neutrinos decaying at different redshifts (both in the radiative and in the not-radiative case). In Section 3 we compare the fluxes derived as explained in Section 2 with the measurements of the background radiation in different bands. First of all, we compare the neutrino flux with the estimates of the X background, trying to put some new constraint on the sterile neutrino mass. Further, we compare our results with the data of the Infrared and Optical background, checking whether sterile neutrinos can significantly contribute to these backgrounds.  In Section 4 we calculate the ionized fraction, at any redshift, due to radiatively and non-radiatively decaying neutrinos and we derive the corresponding Thomson optical depth. Our results are summarized in Section 5.

\section{Light from neutrino decays}
In this Section we calculate the possible contribution of sterile neutrinos decaying at different redshifts to the observed background flux. 
\subsection{Radiative decay channel}
To start, let us consider the case of radiatively decaying neutrinos. Radiative decay is one of the possible outcomes of the following process.
\q\label{eq:eq0}
\nu_{s}\rightarrow{}\nu{}_a+\it{l}+\it{l},
\nq
where $\nu_{s}$ is the sterile neutrino, $\nu{}_a$ is an ordinary active neutrino and {\it{l}} is a lepton.
In particular, this decay is radiative when the two leptons in the previous expression are an electron-positron pair, their annihilation producing a photon. This photon has an energy $E_\gamma{}=(m_{\nu{}s}^2-m_{\nu{}a}^2)/2m_{\nu{}s}$.\\
 Stecker (1980) calculated the flux emitted by sterile neutrinos radiatively decaying at different redshift. Correcting its calculations and updating them to present values of cosmological parameters\footnote{ We adopt the following cosmological parameters: 
Hubble constant $H_0$=72 km s$^{-1}$ Mpc$^{-1}$, $\Omega{}_{0M}\equiv{}\Omega_{DM}+\Omega{}_b$=0.27, 
$\Omega{}_\Lambda{}$=0.73, which are in agreement with the recent WMAP determination (Spergel et al. 2003).}, we find a photon flux (units of cm$^{-2}$ s$^{-1}$ sr$^{-1}$):
\q\label{eq:eq1}
%I(E_{obs})=\frac{1}{4\pi{}}\frac{c}{H_0}\frac{n_s}{\tau{}}\int_0^z e^{-t(z)/\tau{}}\frac{E_0\,{}\delta{}((1+z)E_{obs}-E_0)}{(1+z)\left[(1+z)^3\Omega{}_{0M}+\Omega_{\Lambda{}}\right]^{1/2}}\ud{}{\it z},
%I(E_{obs})=\frac{1}{4\pi{}}\frac{c}{H_0}\frac{n_s}{\tau{}}\int_0^z \frac{e^{-t(z)/\tau{}}\,{}E_0\,{}\delta{}((1+z)E_{obs}-E_0)}{(1+z)\left[(1+z)^3\Omega{}_{0M}+\Omega_{\Lambda{}}\right]^{1/2}}\ud{}{\it z},
I(E_{obs})=\frac{n_s\,{}c}{4\pi{}\,{}H_0\,{}\tau{}}\int_0^z \frac{e^{-t(z)/\tau{}}E_0\,{}\delta{}((1+z)E_{obs}-E_0)}{(1+z)\left[(1+z)^3\Omega{}_{0M}+\Omega_{\Lambda{}}\right]^{1/2}}\ud{}{\it z},
\nq
where $n_s$ and $\tau{}$ are the present number density and the lifetime of sterile neutrinos respectively; $E_0$ and $E_{obs}$ are the emitted and the observed energy of the photon, and $t(z)$ is the time elapsed from the Big Bang to redshift $z$, which can be approximated at high redshift
%, neglecting $\Omega_{\Lambda{}}$, 
as
\q\label{eq:eq2}
t(z)\simeq{}\frac{2}{3}H_0^{-1}\Omega{}_{0M}^{-1/2}(1+z)^{-3/2}
\nq
Since $\delta{}((1+z)E_{obs}-E_0)\neq{}0$
%\Leftrightarrow{}
if and only if $(1+z)=E_0/E_{obs}$, the eq. (\ref{eq:eq1}) becomes (for comparison, see Mass\'o \& Toldr\`a 1999)
\q\label{eq:eq3}
I(E_{obs})=\frac{1}{4\pi{}}\frac{c}{H_0}\frac{n_s}{\tau{}} \frac{e^{-t(E_0/E_{obs})/\tau{}}}{\left[(E_0/E_{obs})^3\Omega{}_{0M}+\Omega_{\Lambda{}}\right]^{1/2}}
\nq
This equation depends on two fundamental, substantially unknown, parameters: the  density $n_s$ and the lifetime $\tau{}$ of sterile neutrinos. A reasonable upper limit of $n_s$ can be obtained by imposing that all the dark matter is composed by sterile neutrinos (Dolgov \& Hansen 2002):
\q\label{eq:eq4}
\frac{n_s}{n_a}=1.2\times{}10^{-2}\left(\frac{\textrm{keV}}{m_{\nu{}s}}\right)\left(\frac{\Omega{}_{DM}}{0.23}\right)\left(\frac{h}{0.72}\right)^2,
\nq
where $n_a$ is the present density of active neutrinos and $m_{\nu{}s}$ is the mass of sterile neutrinos.\\
The lifetime $\tau{}$ can be expressed as a function of $n_s$ and $m_{\nu{}s}$, as follows.
The rate $\Gamma{}$ for the process (\ref{eq:eq0}) is given by (Boehm \& Vogel 1987; Drees \& Wright 2000; Abazajian, Fuller \& Patel 2001a):
\q\label{eq:eq5}
\Gamma{}=\frac{\sin{}^2(2\,{}\theta{})}{768\,{}\pi{}^3}\,{}G_F^2\,{}m_{\nu{}s}^5
\quad{}\simeq{}\frac{\sin{}^2\theta{}}{192\,{}\pi{}^3}\,{}G_F^2\,{}m_{\nu{}s}^5,
\nq
where $\theta{}$ is the mixing angle and $G_F$ the Fermi constant. The branching ratio for radiative decay, that is the fraction of processes of type (\ref{eq:eq0}) leading to the emission of a photon, can be expressed as (Drees \& Wright 2000):
\q\label{eq:eq6}
B=\frac{27\alpha_{em}}{8\pi{}}
\nq
where $\alpha_{em}$ is the fine structure constant. Then, the rate $\Gamma{}_{rad}\equiv{}\tau{}^{-1}$ of radiative neutrino decays is the following.
\q\label{eq:eq7}
\Gamma{}_{rad}=\frac{9\alpha{}_{em}}{512\,{}\pi{}^4}\sin{}^2\theta{}\,{}G_F^2\,{}m_{\nu{}s}^5,
\nq
where $\sin{}^2\theta{}$ can be written as (Dolgov \& Hansen 2002\footnote{In equation (12) of Dolgov \& Hansen 2002 the term $\left(\frac{g}{10.75}\right)^{-1/2}$ should be  $\left(\frac{g}{10.75}\right)^{3/2}$ (S. Hansen private communication).}):
\q\label{eq:eq8}
\sin{}^2\theta{}=2.8\times{}10^{-6}\frac{n_s}{n_a}\frac{c_2^{1/2}}{(1+g_L^2+g_R^2)}\left(\frac{\textrm{keV}}{m_{\nu{}s}}\right)\left(\frac{g}{10.75}\right)^{3/2},
\nq
where $c_2$ is a numerical coefficient depending on the neutrino flavor ($c_2=0.61$ for $\nu{}_e$ and $c_2=0.17$ for $\nu{}_\tau{}$ and $\nu{}_\mu{}$) and $g$ is the number of relativistic degrees of freedom at the time when sterile neutrinos where produced (we adopted $g=10.75$, Viel et al. 2005); finally, $g_R\equiv{}\sin{}^2\theta{}_W$ and $g_L\equiv{}1/2+\sin{}^2\theta{}_W$, where $\sin{}^2\theta{}_W=0.23$ (Dolgov \& Fukugita 1992), $\theta{}_W$ being the weak mixing angle.

\subsection{Pion decay channel}
We can derive the same quantities calculated for radiative decays in the case of pion decay considered by HH. They noticed that, if the sterile neutrino has a mass in the range 140-500 MeV, the dominant decay channel is (Astier et al. 2001)
\q\label{eq:eq9}
\nu_{s}\rightarrow{}\it{l}+\pi{},
\nq
where $\pi{}$ is a pion and {\it l} can be an electron, a positron or a neutrino. We will focus on the case in which {\it l} is an electron. It will have an energy $E_e=(m_{\nu{}s}-m_{\pi{}})\,{}/2$, where $m_{\pi{}}$ is the pion total mass. Then, because the rest mass of the pion is 139.7 MeV, $0<E_e<180$ MeV. As pointed out by HH, an electron in this range of energies has a mean free path to inverse Compton-scattering with CMB photons generally lower than the mean free path to collisionally ionizing hydrogen. Then the most important interaction these electrons will experience is the inverse Compton-scattering with CMB photons. Can these scattered CMB photons significantly contribute to the background radiation? To answer this question we made the following calculations. 
%We can assume, without substantially loosing generality,
As a first approximation, we assume that the electrons produced by neutrino decay follow a power law energy distribution with spectral index $p$ (this assumption takes into account the fact that each electron interacts more than once with CMB photons, gradually degrading its energy). The final spectrum of photons, initially a Planckian, due to inverse Compton-scattering by a power law distribution of electrons is given by (Abramowitz \& Stegun 1965; Rybicki \& Lightman 1979) 
\q\label{eq:eq10}
\frac{\ud{}E}{\ud{}V\,{}\ud{}{\it t}\,{}\ud{}{\it E}\,{}\ud{}\Omega{}}=\frac{1}{4\pi{}}\frac{8\pi{}^2r_0^2\,{}A}{h_{Pl}^3c^2}F(p)\,{}\left(k_BT\right)^{(p+5)/2}\,{}E^{-(p-1)/2}
\nq
where $r_0$ is the classical radius of the electron, $h_{Pl}$ the Planck constant, $k_B$ the Boltzmann constant, $T$ the initial black body temperature of photons, $p$ the spectral index of the electron distribution and $E$ the final photon energy. $F(p)$ is defined by:
\qs
F(p)=2^{(p+3)}\frac{p^2+4\,{}p+11}{(p+3)^2(p+5)(p+1)}\Gamma{\left(\frac{p+5}{2}\right)}\zeta{\left(\frac{p+5}{2}\right)},
\nqs
where $\Gamma{}$ is the Euler's Gamma function and $\zeta{}$ is the Riemann's Zeta function.
$A$ is a normalization factor defined by
\q\label{eq:eq11}
n_e=\int^{\gamma_{max}}_{\gamma_{min}}A\,{}\gamma^{-p}\ud{\gamma{}}
\nq
where $n_e$ and $\gamma{}$ are respectively the density and the Lorentz factor of the electrons ($0<\gamma{}<360$).\\
We included in eq. (\ref{eq:eq10}) the dependences on the redshift and we integrated it over the line-of-sight, obtaining the following equation.
\qa\label{eq:eq12}
I(E_{obs})=\int_0^z\frac{\ud{}E}{\ud{}V\,{}\ud{}{\it t}\,{}\ud{}{\it E}\,{}\ud{}\Omega{}}\frac{\ud{}{\it l}}{\ud{}{\it z}}\ud{}{\it z}\hspace{2.5cm}\nonumber{}\\
\hspace{1.5cm}=\frac{1}{4\pi{}}\frac{8\pi{}^2r_0^2}{h_{Pl}^3c^2}F(p)\,{}\left(k_BT_0\right)^{(p+5)/2}\,{}E_{obs}^{-(p-1)/2}\nonumber{}\\
\times{}\frac{c}{H_0}\int_0^z\frac{A(z)\,{}(1+z)^{-1}}{\left[(1+z)^3\Omega_{0M}+\Omega{}_\Lambda{}\right]^{1/2}}\ud{}{\it z}
\nqa
where $T_0$ is the present temperature of the CMB\footnote{We adopted $T_0=2.275$ K (Spergel et al. 2003).} and $E_{obs}\equiv{}E\,{}(1+z)$ is the final energy of the photon at redshift $z$=0. $A(z)$ can be derived from eq. (\ref{eq:eq11}). In particular we want to express the proper density of electrons $n_e(z)$ as a function of the initial density of sterile neutrinos $n_{s}(z)$ and of their lifetime $\tau{}$. The production rate of electrons is approximately given by $\ud{}{\it n}_e/\ud{}{\it t}={\it n}_{s}({\it z})\exp{[-{\it t}({\it z})/\tau{}]}/\tau{}$. Integrating, we obtain
\q\label{eq:eq13}
n_e(t)=\int^t_0\frac{n_{s}(z)}{\tau{}}e^{-\tilde{t}(z)/\tau{}}\,{}f(\tilde{t})\,{}\ud{}\tilde{{\it t}}
\nq
where $f(t)$ is a function which takes into account the fact that the electrons  have a finite lifetime. In the simplest case, $f(t)$ is a step function
\qs
f(\tilde{t})=\left\{
\begin{array}{l}
1\textrm{ if }(t(z)-t_C(z))<\tilde{t}<t(z)\\
0\textrm{ otherwise}
\end{array}
\right.
\nqs
where $t_C(z)$ is defined as the electron cooling time  
%loose an amount of energy of the order of the mean energy gained by a photon due to Compton-scattering
 due to repeated Compton-scatterings with CMB photons, $t_C(z)\sim{}6\times{}10^4(E_e/100\textrm{ MeV})^{-1}[(1+z)/21]^{-4}\textrm{ yr}$ (HH). Substituting $f(t)$ in eq. (\ref{eq:eq13}), we obtain:
\qa\label{eq:eq14}
n_e(t)=\int^{t(z)}_{(t(z)-t_C(z))}\frac{n_{s}(z)}{\tau{}}e^{-\tilde{t}(z)/\tau{}}\ud{}\tilde{{\it t}}\hspace{0.2cm}\nonumber{}\\
=n_{s}(z)\,{}e^{-t(z)/\tau{}}\left(e^{t_C(z)/\tau{}}-1\right)
\nqa
where $t(z)$ is computed according to eq. (\ref{eq:eq2}).\\
Finally $n_s(z)=n_s\,{}(1+z)^3$ (where $n_s$ is the present sterile neutrino density) can be easily expressed as a function of the present density of active neutrinos $n_a$ and of the lifetime $\tau{}$ and the mass $m_{\nu{}s}$ of sterile neutrinos (HH):
\q\label{eq:eq15}
n_{s}\sim{}\frac{4.8}{(m_{\nu{}s}/m_\pi{})^2-1}n_a\,{}\tau{}^{-1}
%n_{s}\sim{}\frac{4.8\times{}10^{-15}}{(m_{\nu{}s}/m_\pi{})^2-1}n_a\,{}\tau{}^{-1}
\nq
%%%where $\tau{}_{15}\equiv{}\tau{}/10^{15}$ s. The choice of $\tau{}_{15}$ is due to the fact that we want to check the idea of HH that sterile neutrinos can start a reionization phase at redshift $z\sim{}20$. 
Taking into account that $n_a=\frac{3}{11}n_b/\eta{}$ (where $n_b$ is the present baryon density and $\eta{}$ is the present baryon-to-photon ratio\footnote{We used $n_b=2.7\times{}10^{-7}\textrm{ cm}^{-3}$ and $\eta{}=6\times{}10^{-10}$, according to WMAP (Spergel et al. 2003).}), we finally obtain:
%\newpage
\qa\label{eq:eq16}
A(z)=\frac{4.8\,{}(1+z)^3}{\left[(m_{\nu{}s}/m_\pi{})^2-1\right]}\frac{3\,{}n_b}{11\,{}\eta{}}\frac{\tau{}^{-1}}{\int^{\gamma_{max}}_{\gamma_{min}}\gamma^{-p}\ud{\gamma{}}}\nonumber{}\\
%A(z)=\frac{4.8\times{}10^{-15}\,{}(1+z)^3}{\left[(m_{\nu{}s}/m_\pi{})^2-1\right]}\frac{3\,{}n_b}{11\,{}\eta{}}\frac{\tau{}^{-1}}{\int^{\gamma_{max}}_{\gamma_{min}}\gamma^{-p}\ud{\gamma{}}}\nonumber{}\\
\times{}\,{}e^{-t(z)/\tau{}}\left(e^{t_C(z)/\tau{}}-1\right), 
\nqa
which can be substituted into eq. (\ref{eq:eq12}).
%%%Therefore we can write the eq. \ref{eq:eq12} as
%%%\qa\label{eq:eq17}
%%%I(E_{obs})=\frac{4.8}{\left[(m_{\nu{}s}/m_\pi{})^2-1\right]}\frac{3\,{}n_b}{11\,{}\eta{}}\frac{\tau{}^{-1}}{\int^{\gamma_{max}}_{\gamma_{min}}\gamma^{-p}\ud{\gamma{}}}\nonumber{}\\
%%%\times{}\frac{1}{4\pi{}}\frac{8\pi{}^2r_0^2}{h_{Pl}^3c^2}F(p)\,{}\left(k_BT_0\right)^{(p+5)/2}\,{}E_{obs}^{-(p-1)/2}\nonumber{}\\
%%%\times{}\frac{c}{H_0}\int_0^z\frac{(1+z)^2}{\left[(1+z)^3\Omega_{0M}+\Omega{}_\Lambda{}\right]^{1/2}}\nonumber{}\hspace{1.2cm}\\
%%%\times{}\,{} e^{-\frac{2}{3}H_0^{-1}\Omega{}_{0M}^{-1/2}(1+z)^{-3/2}}\left(e^{t_C(z)/\tau{}}-1\right)\ud{}{\it z}
%%%\nqa

\section{Background radiation}
Now we have all the equations we need to estimate the contribution of sterile neutrinos to the background radiation. Using  eqs. (\ref{eq:eq3}) and (\ref{eq:eq12}) we can derive the observed flux respectively due to radiative and not-radiative neutrino decays.  
%under the hypothesis that the absorption of photons is negligible. 
In our calculations, we assume that all the photons produced (or Compton-scattered) by neutrinos decaying at $z>1000$ are thermalized by CMB photons, and then are not visible. We assume also, for simplification, that all the photons  produced (or Compton-scattered) by neutrinos decaying at $z<1000$ are not absorbed or scattered. For this reason, our calculation represents an upper limit for the optical and infrared background; however it should give a good estimate of the X-ray flux.
\subsection{Soft X-ray background }
%%%%%%\subsubsection{Constraints on the SXRB}
A recent estimate of the soft X-ray background (SXRB), in the energy range 0.5-2.0 keV, is provided by M03. Combining 10 different measurements reported in the literature, they found a flux of $2.47\pm{}0.11\,{}\times{}10^{-8}$ erg s$^{-1}$ cm$^{-2}$ sr$^{-1}$, which must be lowered to about 1.48$\times{}10^{-9}$ erg s$^{-1}$ cm$^{-2}$ sr$^{-1}$, if the mean contribution of both point and diffuse unresolved sources is subtracted at the level estimated by M03. D04 derive the slightly lower value $1.15\pm{}1.64\,{}\times{}10^{-9}$ erg s$^{-1}$ cm$^{-2}$ sr$^{-1}$, as 
they subtract a further 1.0-1.7\% contribution by the diffuse component due to Thomson-scattered point source radiation 
(Soltan 2003).
%%%%%%Then, from the analysis of Dijkstra et al, the flux due to unaccounted sources is $1.15\pm{}1.64\,{}\times{}10^{-9}$ erg s$^{-1}$ cm$^{-2}$ sr$^{-1}$, derived subtracting to the observed flux both the mean contribution of point and diffuse unresolved sources (at the level estimated by Moretti et al.) and the mean scattered flux from free electrons (as calculated by Soltan). 
D04 also suggest a maximum possible value $\sim{}4.04\,{}\times{}10^{-9}$ erg s$^{-1}$ cm$^{-2}$ sr$^{-1}$, by subtracting the lower limit of unresolved sources and of Thomson scattered flux.
D04             further notice that the theoretically expected amount of X-ray emission in the soft band from thermal emission by gas in clusters (Wu \& Xue 2001) should represent $\sim{}$9\% of the total SXRB, considerably higher than the 6\% estimated by M03.            However, D04             do not include in their estimate of the background this probable additional contribution. Otherwise, the  flux due to unaccounted sources should be lowered to $\sim{}4.0\times{}10^{-10}$ erg s$^{-1}$ cm$^{-2}$ sr$^{-1}$.\\
More recently, B04 investigated the X-ray number counts in the Chandra Deep Fields (CDFs). Adopting  for the total SXRB from 0.5 to 2 keV the value suggested by M03            (i.e. $2.47\pm{}0.11\,{}\times{}10^{-8}$ erg s$^{-1}$ cm$^{-2}$ sr$^{-1}$), they found that its resolved fraction is $89.5^{+5.9}_{-5.7}$\%. The resolved sources are predominantly AGNs ($\sim{}83$\%) and star forming galaxies ($\sim{}3$\%).  This means that unresolved sources produce a flux $2.59_{-1.46}^{+1.41}\times{}10^{-9}$ erg s$^{-1}$ cm$^{-2}$ sr$^{-1}$.
%%%%%%%%%%%%%%%%%%%%%%%%%%%%%%%%%%% FIGURE 1 %%%%%%%%%%%%%%%%%%%%%%%%%%%%%%%%%%
\begin{figure}
\center{{
\epsfig{figure=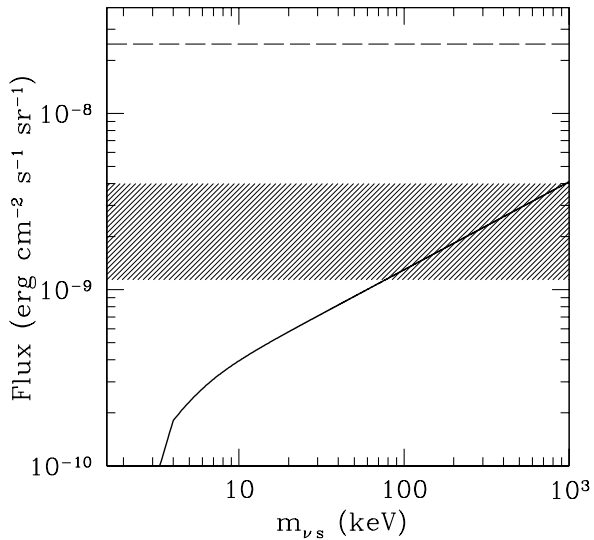,height=8cm}
}}
\caption{\label{fig:fig1} 
Integrated flux between 0.5 and 2 keV due to radiatively decaying neutrinos as a function of the neutrino mass ({\it solid line}). The horizontal {\it dashed line} represents the total measured background (M03). The shaded area indicates the flux due to unresolved sources (B04).           
} 
\end{figure}
%%%%%%%%%%%%%%%%%%%%%%%%%%%%%%%%%%%%%%%%%%%%%%%%%%%%%%%%%%%%%%%%%%%%%%%%%%%%%%%

These strong limits on the SXRB are important to constraint the mass of sterile neutrinos.
We base our analysis on the paper by B04; in addition we have checked the values derived by M03            and by D04.
The latter are consistent with those reported by B04 and give only slightly different constraints on the 
sterile neutrino mass.
%%%%%%\subsubsection{Sterile neutrinos as sources of SXRB}
In Fig.~1 we report the integrated flux between 0.5 and 2 keV due to radiative neutrino decays for different neutrino masses. To calculate this flux we derived the neutrino comoving density from eq. (\ref{eq:eq4}), i.e. assuming that all the dark matter is due to sterile neutrinos. Given the mass and the density, the lifetime is completely determined by eqs. (\ref{eq:eq7}) and (\ref{eq:eq8}). The flux due to sterile neutrinos is compared with the total flux predicted by M03                   and with the flux due to unresolved sources (B04).                     
 %We assume that all the photons produced by neutrinos decaying at $z>1000$ are thermalized by CMB photons, and then are not visible. 
As we can see from Fig. 1, 
%radiatively decaying sterile neutrinos never exceed the flux  
%$2.47\,{}\times{}10^{-8}$ erg s$^{-1}$ cm$^{-2}$ sr$^{-1}$ derived by M03.                     
%We find that
 only neutrinos with masses $\lesssim{}$950 keV do not exceed the maximum value of the flux due to unresolved sources (4.00$\times{}10^{-9}$ erg s$^{-1}$ cm$^{-2}$ sr$^{-1}$, B04); the same conclusion can be drawn by using D04 data.
%This is a strong constraint, if combined with the lower limit for the sterile neutrino mass $\gtrsim{}$2 keV derived by 
%Viel et al. 2005. We must be aware of two uncertainties in our calculations. First, we made the assumption that all the 
%dark matter is composed by sterile neutrinos. In addition, we have not taken into account the clustering of matter at low 
%redshift, which can lead to an overestimate of the flux due to neutrinos (Abazajian et al. 2001b).
%%%%%%%%%%%%%%%%%%%%%%%%%%%%%%%%%%% FIGURE 2 %%%%%%%%%%%%%%%%%%%%%%%%%%%%%%%%%%
\begin{figure}
\center{{
\epsfig{figure=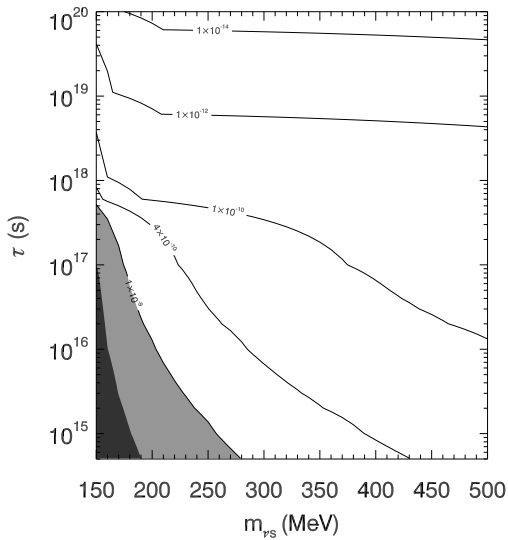,height=8cm}
}}
\caption{\label{fig:fig2} 
Isocontours of integrated X-ray flux  in the band 0.5-2 keV (units of erg cm$^{-2}$ s$^{-1}$ sr$^{-1}$) due to pion-decaying neutrinos  as a function of the neutrino mass and lifetime. The light gray area indicates flux due to unresolved sources (B04). The dark gray area indicates the ''forbidden'' region in which the flux exceeds the  maximum background flux due to unaccounted sources (B04; M03).    
} 
\end{figure}
%%%%%%%%%%%%%%%%%%%%%%%%%%%%%%%%%%%%%%%%%%%%%%%%%%%%%%%%%%%%%%%%%%%%%%%%%%%%%%%

We repeat the same calculation for heavy sterile neutrinos ($140<m_{\nu{}s}<500$ MeV) decaying into pions and electrons. In this case we consider the flux due to CMB photons Compton-scattered by the electrons coming from the neutrino decay (eq. (\ref{eq:eq12}), assuming $p$=1). Fig.~2 shows the isocontours of the integrated flux between 0.5 and 2 keV due to Compton-scattered CMB photons, as a function of the progenitor neutrino mass and lifetime. 
%We fixed the lifetime at $\tau{}=10^{15}$ s. 
As we can see from the eq. (\ref{eq:eq15}), the density of neutrinos depends strongly on their mass and on the lifetime. In particular, less massive neutrinos have much higher comoving density. For this reason, and because of the largely different mass range,  the observed flux has an opposite behavior with respect to radiatively decaying neutrinos. In this case the soft X-ray flux decreases increasing the neutrino mass. 
The flux produced by 150-500 MeV neutrinos does not violate the SXRB limit (B04), if the lifetime is $\geq{}10^{17}$~s. 
For shorter lifetimes, neutrinos with mass lower than 190 MeV do violate the constraint imposed 
by unresolved sources, according to both M03/D04 and B04 estimates.

%%%%%%%%%%%%%%%%%%%%%%%%%%%%%%%%%%% FIGURE 3 %%%%%%%%%%%%%%%%%%%%%%%%%%%%%%%%%%
%%%%%\begin{figure}
%%%%%\center{{
%%%%%\epsfig{figure=SXB_decpion_tau.eps,height=8cm}
%%%%%}}
%%%%%\caption{\label{fig:fig3} 
%%%%%Integrated flux between 0.5 and 2 keV due to CMB photons Compton scattered by electrons coming from neutrino decays, as a function of the neutrino lifetime ({\it solid line}). The neutrino mass is fixed ($m_{\nu{}s}=250$ MeV). The horizontal lines are the same as in Fig.~1.
%%%%%} 
%%%%%\end{figure}
%%%%%%%%%%%%%%%%%%%%%%%%%%%%%%%%%%%%%%%%%%%%%%%%%%%%%%%%%%%%%%%%%%%%%%%%%%%%%%%

\subsection{Hard X-ray background}
%%%%%%\subsubsection{Constraints on the HXRB}
%%%%%Gruber (1992) found the form of the X-ray background, in the range 3-60 keV, to be
%%%%%\q\label{eq:gruber}
%%%%%I(E_{obs})\lesssim{}7.9\,{}\left(\frac{E_{obs}}{\textrm{keV}}\right)^{-0.29}\,{}e^{-(E_{obs}/41\textrm{ keV})}\,{}\textrm{cm}^{-2}\textrm{ s}^{-1}\textrm{ sr}^{-1}
%%%%%\nq
%%%%%More recent measurements (Hornschmeier et al. 2001; Tozzi et al. 2001) have resolved sources which contribute from 60\% to 90\% of the previously unresolved background, indicating that the value found by Gruber must be substantially lowered. 
M03                   estimated the total background flux in the hard band (2-10 keV, HXRB) to be $6.63\pm{}0.36\times{}10^{-8}$ erg cm$^{-2}$ s$^{-1}$ sr$^{-1}$. They also show that the resolved fraction of this background is $88.8^{+7.8}_{-6.6}$\%.Then only the 11.2\% of the hard X-ray background can be due to unaccounted sources (like sterile neutrinos).\\
B04 studied the HXRB flux from 2 to 8 keV, using the X-ray number counts in the CDFs. They adopted a total HXRB flux $5.88\pm{}0.36\times{}10^{-8}$ erg cm$^{-2}$ s$^{-1}$ sr$^{-1}$ and found that the resolved fraction is $92.6^{+6.6}_{-6.3}$, dominated by AGNs ($\sim{}95$\%). This means that the unresolved flux is only $4.35_{-3.88}^{+3.70}\times{}10^{-9}$ erg cm$^{-2}$ s$^{-1}$ sr$^{-1}$.
 
%%%%%%\subsubsection{Sterile neutrinos as sources of HXRB}

%%%%%%%%%%%%%%%%%%%%%%%%%%%%%%%%%%% FIGURE 3 %%%%%%%%%%%%%%%%%%%%%%%%%%%%%%%%%%
\begin{figure}
\center{{
\epsfig{figure=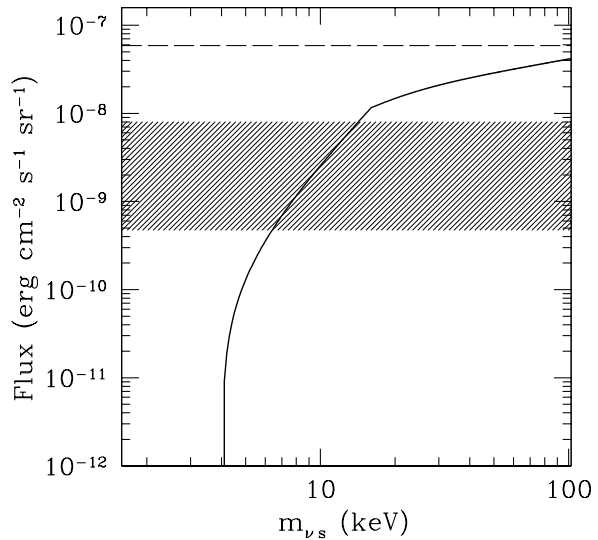,height=8cm}
}}
\caption{\label{fig:fig3} 
Integrated X-ray flux in the range 2-8 keV due to radiatively decaying neutrinos as a function of the neutrino mass ({\it solid line}). The horizontal {\it dashed line} indicates the total background in the same band; the 
shaded area indicates the flux due to unresolved sources (B04).}
%%%%%Horizontal lines indicate the total level of the background between 2 and 10 keV ({\it solid line}) and the fraction due to unaccounted sources ({\it dotted line}), as evaluated by Moretti et al. (2003).
 
\end{figure}
%%%%%%%%%%%%%%%%%%%%%%%%%%%%%%%%%%%%%%%%%%%%%%%%%%%%%%%%%%%%%%%%%%%%%%%%%%%%%%%

We compared the hard X-ray flux due to sterile neutrinos with the background level estimated by B04, and we checked for consistency with M03.                     
In particular, for radiatively decaying sterile neutrinos we found that the integrated flux in the range 2-8 keV is lower than the background flux due to unresolved sources, as estimated by B04, only if $m_{\nu{}s}\lesssim{}14$ keV (Fig.~3). If we consider  the unresolved flux in the range 2-10 keV estimated by M03, we find a similar constraint for the neutrino mass: $m_{\nu{}s}\lesssim{}16$ keV.
This is a strong constraint, if combined with the lower limit for the sterile neutrino mass $\gtrsim{}$2 keV derived by 
Viel et al. 2005. We must be aware of two uncertainties in our calculations. First, we made the assumption that all the 
dark matter is composed by sterile neutrinos. In addition, we have not taken into account the clustering of matter at low 
redshift, which can lead to an overestimate of the flux due to neutrinos (Abazajian et al. 2001b).
%%%%%Using the model for the X-ray background described by Gruber in the range 1-60 keV, we find that acceptable sterile neutrino masses are $m_{\nu{}s}\lesssim{}8.5$ keV. If we take into account that only the 10-40\% of Gruber's background model (Tozzi et al. 2001) comes from unresolved sources, we can accept only neutrino masses respectively $m_{\nu{}s}<3$ keV, or $m_{\nu{}s}<6$ keV (Fig.~5), consistent with the findings of Abazajan, Fuller \& Tucker (2001b).

%%%%%%%%%%%%%%%%%%%%%%%%%%%%%%%%%%% FIGURE 4 %%%%%%%%%%%%%%%%%%%%%%%%%%%%%%%%%%
%%%%%\begin{figure}
%%%%%\center{{
%%%%%\epsfig{figure=HXB_pion.eps,height=8cm}
%%%%%}}
%%%%%\caption{\label{fig:fig4} 
%%%%%Integrated flux between 2 and 10 keV due to CMB photons Compton scattered by electrons produced by heavy sterile neutrinos as a function of the neutrino mass ({\it solid line}). Horizontal lines indicate the total level of the background between 2 and 10 keV ({\it solid line}) and the fraction due to unaccounted sources ({\it dotted line}), as evaluated by Moretti et al. (2003).
%%%%%} 
%%%%%\end{figure}
%%%%%%%%%%%%%%%%%%%%%%%%%%%%%%%%%%%%%%%%%%%%%%%%%%%%%%%%%%%%%%%%%%%%%%%%%%%%%%%
%%%%%%%%%%%%%%%%%%%%%%%%%%%%%%%%%%% FIGURE 4 %%%%%%%%%%%%%%%%%%%%%%%%%%%%%%%%%%
\begin{figure}
\center{{
\epsfig{figure=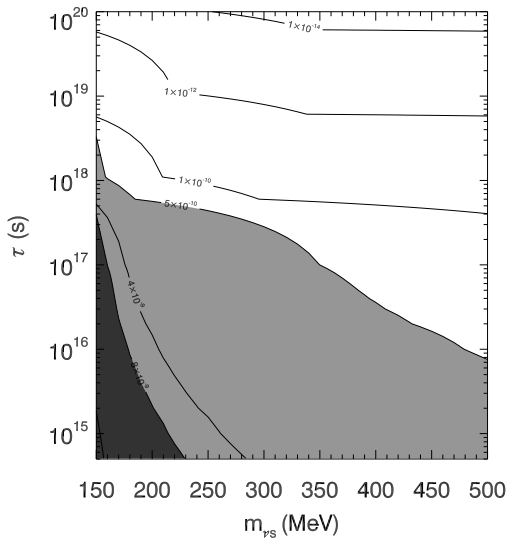,height=8cm}
}}
\caption{\label{fig:fig4} 
Isocontours of the integrated X-ray flux in the band 2-8 keV (units of erg cm$^{-2}$ s$^{-1}$ sr$^{-1}$)  due to pion-decaying neutrinos as a function of the neutrino mass and lifetime. The light gray area indicates the flux due to unresolved sources (B04). The dark gray area indicates the ''forbidden'' in which the flux  exceeds the maximum background flux due to unresolved sources (B04).}              
 
\end{figure}
%%%%%%%%%%%%%%%%%%%%%%%%%%%%%%%%%%%%%%%%%%%%%%%%%%%%%%%%%%%%%%%%%%%%%%%%%%%%%%%

On the other hand, the integrated flux (in the range 2-8 keV) of CMB photons scattered by heavy pion-decay neutrinos 
is lower than the upper limit of the background due to unresolved sources (B04) for all the neutrino masses from 150 to 500 MeV, if the lifetime is longer than $\sim{}4\times{}10^{17}$~s (Fig.~4). We obtain the same result by considering (in the range 2-10 keV) the flux due to unaccounted sources as estimated by M03. For shorter lifetimes, the minimum allowed mass increases. For example, if $\tau{}=10^{15}$ s, neutrinos with masses lower than $\sim{}$215 MeV exceed the upper limit of the background due to unresolved sources as derived by B04; whereas, using the estimate of M03           , we find $\sim{}$200 MeV. 
%%%%%On the other hand, the integrated flux (in the range 2-10 keV) of CMB photons scattered by heavy neutrinos decaying into pions and electrons is less than the value predicted by Moretti et al. only for neutrino masses higher than 240 MeV (Fig.~4). 
%%%%%If we adopt the model of Gruber, only neutrinos with masses $m_{\nu{}s}\gtrsim{}250$ MeV do not exceed the total background and only neutrinos with masses $m_{\nu{}s}\gtrsim{}350$ MeV do not exceed the background due to unaccounted sources in the range 3-60 keV (Fig.~7). 

In summary, using the HXRB we put much stronger constraints on the mass of radiatively decaying neutrinos ($m_{\nu{}s}\lesssim{}14$ keV) than using the SXRB. Instead, in the case of heavy sterile neutrinos decaying into pions, all the masses from 150 to 500 MeV are allowed, provided a lifetime longer than $4\times{}10^{17}$ s. 
%%%%%%%For shorter lifetimes, the minimum allowed mass increases. For example, if $\tau{}=10^{15}$ s, $215\lesssim{}m_{\nu{}s}\lesssim{}500$ MeV).

%%%%%%%Again uncertainties come from the assumption that the background due to sterile neutrinos is isotropic. This assumption is not completely corrected at low redshift, because of dark matter clustering (Abazajian et al. 2001b). 

To compare our results with those obtained by Abazajian et al. (2001b), we have repeated this analysis using Gruber's model of HXRB (1992). We found that, assuming this model of HXRB, the mass of radiatively decaying neutrinos must be $m_{\nu{}s}\lesssim{}3$ keV, in agreement with Abazajian et al. (2001b). However, in the following we will use the constraints derived using the estimates by B04 and M03, because they come from more recent data and do not require any assumption about the shape of the HXRB due to unresolved sources.

\subsection{Optical and near-infrared (NIR) background}
%%%%%%\subsubsection{Constraints on the NIR/optical background}
A number of factors, such as interplanetary dust scattered sunlight (zodiacal light), terrestrial airglow, dust-scattered Galactic starlight (diffuse Galactic light), make very difficult to obtain a reliable measurement of the NIR and especially of the optical background light. Here we briefly summarize the most recent measurements of optical and NIR background.
Deep optical and NIR galaxy counts give an estimate of the EBL fraction
coming from normal galaxies. Madau \& Pozzetti (2000) derived the contribution of 
known galaxies in the $UBVIJHK$ bands from the $Southern\,{}Hubble\,{}Deep\,{}Field$ 
imaging survey. In particular for the U, V, B and I bands (corresponding to the wavelengths $\lambda{}$ = 3600, 4500, 6700 and 8100 \AA) they found a mean flux respectively $2.87_{-0.42}^{+0.58}$, $4.57_{-0.47}^{+0.73}$, $6.74_{-0.94}^{+1.25}$ and $8.04_{-0.92}^{+1.62}$ in units of  10$^{-6}$ erg s$^{-1}$ cm$^{-2}$ sr$^{-1}$. Following a different approach, Bernstein et al. (2002) measured the mean flux of the optical EBL at 3000, 5500, and 8000 \AA, using the Wide Field Planetary Camera 2 (WFPC2) and the Faint Object Spectrograph, both on board the {\it Hubble Space Telescope}, combined with the du Pont 2.5 m Telescope at the Las Campanas Observatory. They found for these three band a mean flux of the EBL respectively $12.0_{-6.3}^{+17.7}$, $14.9_{-10.5}^{+19.3}$, and $17.6_{-12.8}^{+22.4}$ in units of 10$^{-6}$ erg s$^{-1}$ cm$^{-2}$ sr$^{-1}$, considerably higher than the contribution of the galaxy counts alone. 

The available NIR background data come from the Diffuse Infrared Background Experiment (DIRBE) on board of the Cosmic Background Explorer (COBE) and from the Near InfraRed Spectrometer (NIRS) on board of the InfraRed Telescope in Space (IRTS). A summary of the DIRBE results can
be found in Hauser et al. (1998). Matsumoto et al. (2000) made a preliminary analysis of the NIRS data, estimating the NIRB on the basis of the 5 NIRS observation days unperturbed by atmospheric, lunar and nuclear radiation effects.  
Both the DIRBE and the NIRS data show an excess in the NIRB with respect to galaxy counts. The amount of this excess depends on the model used to subtract the contribution of the zodiacal light from the measurements. In fact, there are two different models of zodiacal light, that described in Kelsall et al. (1998) and that presented by Wright 1997 (see also Wright 1998 and Wright \& Reese 2000). 
Fig.~5 presents a summary of all the data we have considered.
%available data.

%%%%%%%%%%%%%%%%%%%%%%%%%%%%%%%%%%% FIGURE 5 %%%%%%%%%%%%%%%%%%%%%%%%%%%%%%%%%%
\begin{figure}
\center{{
\epsfig{figure=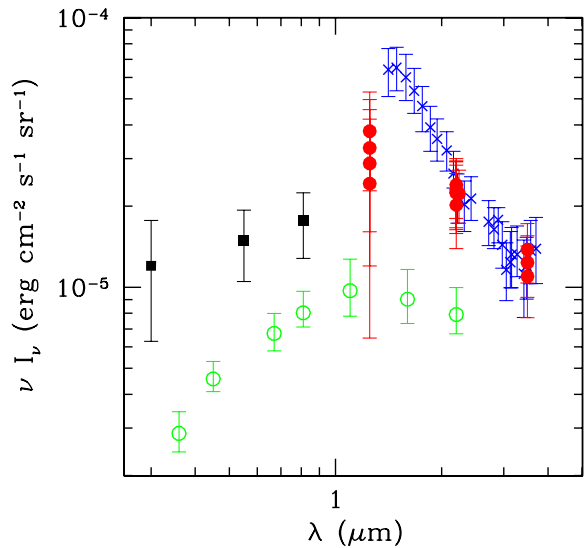,height=8cm}
}}
\caption{\label{fig:fig5} 
Measurements of the optical and NIR background: Bernstein et al. ({\it filled squares}); Madau \& Pozzetti ({\it open  circles}); Matsumoto et al. ({\it crosses}); DIRBE data reduced adopting the Wright model of zodiacal light ({\it filled circles}).
} 
\end{figure}
%%%%%%%%%%%%%%%%%%%%%%%%%%%%%%%%%%%%%%%%%%%%%%%%%%%%%%%%%%%%%%%%%%%%%%%%%%%%%%%

The measurements of the NIR and, with less evidence, of the optical background, seem to indicate the presence of an excess with respect to galaxy counts. The most plausible explanation of the NIR excess 
 is that the extragalactic background light in this wavelength range is due to the redshifted UV and optical light emitted by Pop~III stars (Bond, Carr \&{} Hogan 1986; Santos, Bromm \&{} Kamionkowski 2002; Salvaterra \&{} Ferrara 2003). In particular Salvaterra \&{} Ferrara (2003) developed a model of the NIR background which, accounting for the most recent predictions of  Pop~III stellar spectra (Schaerer 2002) and IMF, nebular emission (i.e. the radiation coming from the nebula surrounding the star), and L$_{y\alpha{}}$ photons scattered by the intergalactic medium, is able to fit the NIRS data (Matsumoto et al. 2000) and the DIRBE data with both the methods of zodiacal light subtraction. 
%Their best fit predicts, for a star formation efficiency $f_{\ast{}}=0.1-0.5$, depending on the adopted IMF, a transition from (very massive)
%Pop~III to Pop~II stars occurring at $z\approx 9$. 
This model is supported also by the analysis of the infrared background fluctuations performed by Magliocchetti et al. (2003).

In principle, sterile neutrinos could be an additional source of optical and infrared light (Stecker 1980; Biller et al. 1998). In this Section we want to check this hypothesis, calculating the contribution of sterile neutrinos (in the range of masses and/or lifetimes allowed by the constraints derived from the HXRB) to the optical and infrared background and comparing it with observational data.

%%%%%%\subsubsection{Are sterile neutrinos sources of the NIR and optical background excess?}

In Fig.~6 we plot the contribution to the background light by radiatively decaying neutrinos for three different allowed masses (respectively 2, 6 and 14 keV). We derived their density by assuming that all the dark matter is composed by sterile neutrinos. The lifetime is uniquely determined by the neutrino density and mass. We conclude that the flux due to radiatively decaying neutrinos in the optical and infrared background is completely negligible, being at least 8 order of magnitudes lower than the observational data (shown in Fig.~5) even in the most favorable hypothesis. We notice that the cut-off at low energy is due to the assumption that photons emitted at $z>1000$ are completely thermalized. The high energy cutoff simply follows from the fact that it has to be $z\geq{}0$.
 
%%%%%%%%%%%%%%%%%%%%%%%%%%%%%%%%%%% FIGURE 6 %%%%%%%%%%%%%%%%%%%%%%%%%%%%%%%%%%
\begin{figure}
\center{{
\epsfig{figure=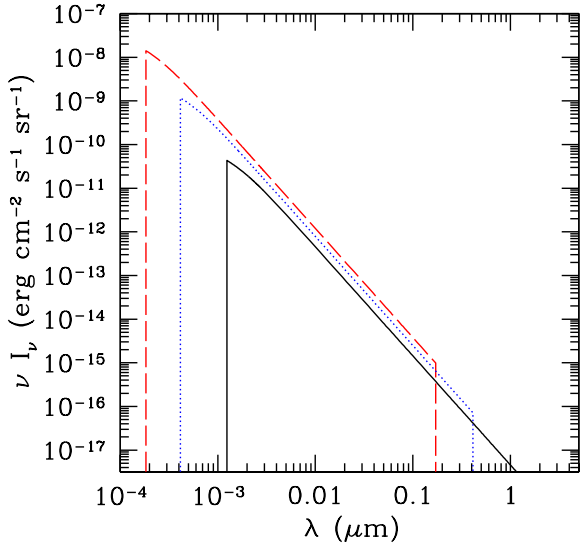,height=8cm}
}}
\caption{\label{fig:fig6} 
Background due to radiatively decaying neutrinos with mass 14~keV ({\it dashed line}), 6~keV ({\it dotted}) and 2 keV ({\it solid}). 
} 
\end{figure}
%%%%%%%%%%%%%%%%%%%%%%%%%%%%%%%%%%%%%%%%%%%%%%%%%%%%%%%%%%%%%%%%%%%%%%%%%%%%%%%
%%%%%%%%%%%%%%%%%%%%%%%%%%%%%%%%%%% FIGURE 7 %%%%%%%%%%%%%%%%%%%%%%%%%%%%%%%%%%
\begin{figure}
\center{{
\epsfig{figure=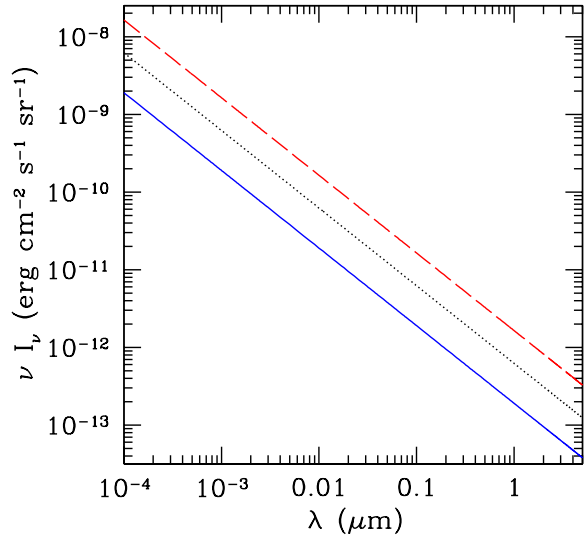,height=8cm}
}}
\caption{\label{fig:fig7} 
Background due to CMB photons Compton-scattered by electrons produced by sterile neutrinos of masses 215~MeV ({\it dashed line}), 300~MeV ({\it dotted}) and 500 MeV ({\it solid}). Neutrino lifetime is $\tau{}=10^{15}$ s; spectral index of electrons $p$=1. 
} 
\end{figure}
%%%%%%%%%%%%%%%%%%%%%%%%%%%%%%%%%%%%%%%%%%%%%%%%%%%%%%%%%%%%%%%%%%%%%%%%%%%%%%%

Fig.~7 shows that also the optical and infrared background due to CMB photons Compton-scattered by electrons which are the product of the decay of massive sterile neutrinos\footnote{We assumed spectral index of the electron distribution $p$=1, which maximizes the optical and infrared flux.}
is negligible with respect to the observed background, plotted in Fig.~5.
%%% (assuming spectral index of the electron distribution $p$=1, which maximizes the optical and infrared flux).
%, at least if we assume a neutrino lifetime ($\tau{}\gtrsim{}10^{15}$ s) that does not violate the SXRB (Fig.~3).

From our analysis we have to conclude that sterile neutrinos cannot significantly contribute to optical and infrared background. Discarding the hypothesis of sterile neutrinos on the basis of our results, the only models which can explain the excess observed in the NIR background are those involving the contribution of Population III stars (Salvaterra \& Ferrara 2003; Madau \& Silk 2005) and/or of high redshift miniquasars (Madau et al. 2004; Cooray \& Yoshida 2004). 
%%%%%Also the hypothesis of miniquasars, that is of intermediate mass black holes produced as final stage of the life of very massive Population III stars, must probably be discarded, because their contribution to the NIR background is negligible with respect to stars and since they can easily exceed the X-ray background, except for a very low accretion efficiency (Salvaterra et al. in preparation). An alternative possible candidate for the production of the optical and NIR background is a population of normal galaxies, as described in the semi-analytic models of Primack et al. (2000). But these galaxies cannot reproduce the fluctuations measured in the NIR background, as, on the contrary, Population III stars do (Magliocchetti et al. 2003). Therefore, Population III and old Population II stars seem to be the only plausible candidates for the production of the NIR background, even if also this model presents some not negligible problem (Madau \& Silk 2005; Schneider et al. in preparation).
\section{Impact on reionization}
As briefly discussed in the Introduction, the most recent measurements of the Thomson optical depth (Springel et al. 2003; Kogut et al. 2003; Fan et al. 2003) suggest a complex reionization history, difficult to model without invoking the contribution of ''exotic'' sources (like very massive Population III stars and miniquasars). Sterile neutrinos have been also proposed as ionization sources, both in the case of radiative decays (Sciama 1982), and, more recently, of pion-decays (HH). In this Section we reexamine this hypothesis, using the calculations presented in Section 2. 
\subsection{Radiative decay channel}
Let us consider first the possible role of radiatively decaying sterile neutrinos. To derive the Thomson optical depth provided by this process, we calculate the hydrogen ionization fraction, $x$, due to photons emitted by neutrino decays. Under the hypothesis of ionization equilibrium, this can be calculated (Osterbrock 1988) as:
\q\label{eq:eq18}
(1-x)\,{}n_H\int_{E_{th}}^{\infty{}}4\pi{}\frac{\ud{}N}{\ud{}{\it E}\,{}\ud{}A\,{}\ud{}{\it t}}\sigma{}_E\,{}{\ud{}}{\it E}=x^2\,{}n_H^2\alpha{}(T),
\nq
where $n_H=n_{H0}+n_{H+}$ is the total hydrogen number density ($n_{H0}$ and $n_{H+}$ being the neutral and ionized hydrogen number density respectively);
%%% considering both neutral ($n_{H0}$) and ionized atoms ($n_{H+}$); 
$x\equiv{}n_{H+}/n_{H}$ is the ionization fraction; $E_{th}=13.6$ eV, $\ud{}N/\ud{}{\it E}\,{}\ud{}A\,{}\ud{}{\it t}$ is the photon flux (units of cm$^{-2}$ s$^{-1}$ sr$^{-1}$ erg$^{-1}$), $\alpha{}(T)=4.18\times{}10^{-13}(T/10^4\textrm{K})^{-0.726}$ cm$^3$ s$^{-1}$ is the recombination coefficient; 
%%%(we assumed $\alpha{}(T=10000\,{}\textrm{K})=4.18\times{}10^{-13}$ cm$^3$ s$^{-1}$, Osterbrock 1988). 
$\sigma{}_E$ is the photo-ionization cross section of hydrogen atoms. 
After some calculations reported in Appendix~A,  in the case of radiatively decaying sterile neutrinos eq. (\ref{eq:eq18}) can be expressed as: 
\qa\label{eq:eq26}
\frac{x^2(z)}{(1-x(z))}=\frac{n_s\,{}(1+z)}{\tau{}\,{}\alpha{}(T)\,{}n_H(0)}\,{}\frac{c}{E_0\,{}H_0\,{}[\Omega_{0M}(1+z)^3+\Omega{}_{\Lambda{}}]^{1/2}}
\nonumber{}\\
\times{}\int_{E_{th}/(1+z)}^{E_0/(1+z)}\frac{\sigma{}_{E(z)}\,{}e^{-t(E_0/E_{obs})/\tau{}}\,{}e^{-\tau{}_{abs}(E_0/E_{obs})}}{\left[\left(\frac{E_0}{E_{obs}}\right)^3\frac{\Omega{}_{0M}}{[\Omega_{0M}(1+z)^3+\Omega{}_{\Lambda{}}]}+\Omega_{\Lambda{}}\right]^{1/2}}\,{}{\ud{}}E_{obs}
\nqa
where $\tau{}_{abs}$ is defined by eq. (\ref{eq:eqtau}).

So far, we have neglected the possibility that the ionizing photon produces a secondary electron which is sufficiently energetic to collisionally ionize additional hydrogen atoms. This is an important effect if if $E(z)\gg{}E_{th}$. To take 
it into account, we modify the cross section by introducing a factor $[1+\phi{}(x(z))\,{}E(z)/E_{th}]$, where  
(Shull \& van Steenberg 1985) $\phi{}(x)={\mathcal{C}}\,{}(1-x^a)^b$; for hydrogen ionization ${\mathcal{C}}=0.3908$, $a=0.4092$ and $b=1.7592$.
Adding this term in  eq. (\ref{eq:eq26}) we finally obtain:
\qa\label{eq:eq27}
\frac{x^2(z)}{(1-x(z))}=\frac{n_s\,{}(1+z)}{\tau{}\,{}\alpha{}(T)\,{}n_H(0)}\,{}\frac{c}{E_0\,{}H_0\,{}[\Omega_{0M}(1+z)^3+\Omega{}_{\Lambda{}}]^{1/2}}
\nonumber{}\\
\times{}\int_{E_{th}/(1+z)}^{E_0/(1+z)}\frac{e^{-t(E_0/E_{obs})/\tau{}}\,{}e^{-\tau{}_{abs}(E_0/E_{obs})}}{\left[\left(\frac{E_0}{E_{obs}}\right)^3\frac{\Omega{}_{0M}}{[\Omega_{0M}(1+z)^3+\Omega{}_{\Lambda{}}]}+\Omega_{\Lambda{}}\right]^{1/2}}\nonumber{}\\
\times{}\quad{}\sigma{}_{E(z)}\,{}\left[1+\phi{}(x(z))\frac{E_{obs}(1+z)}{E_{th}}\right]\,{}{\ud{}}E_{obs}
\nqa
%In this equation we have a slight dependence of the right term on the unknown $x(z)$ ($\phi{}$ and  $\tau{}_{abs}(E_0/E_{obs})$ being functions of $x(z)$). 
We solve this equation iteratively, evaluating $\phi{}$ and $\tau{}(E_0/E_{obs})$ at $z_{i-1}$, where $(z_i-z_{i-1})\lesssim{}\epsilon{}$, with $\epsilon{}$ suitably small (in our calculations $\epsilon{}=0.1$ gives a good result). 

Finally, we have to include the relic ionization fraction remaining after recombination. At high redshift this fraction can be higher than that due to sterile neutrinos. Therefore, to evaluate $\phi{}(x)$ and $\tau{}_{abs}(E_0/E_{obs})$, we take a value  $x(z)=max[x_{\nu{}}(z),x_{rel}(z)]$, where  $x_{\nu{}}(z)$ is the ionization fraction due to sterile neutrino decays and $x_{rel}(z)$ is the residual ionization fraction. For $x_{\nu{}}(z)$ we use the value found by solving eq. (\ref{eq:eq27}). For $x_{rel}(z)$ we adopt the values derived by Tegmark, Silk, Rees Blanchard, Abel \& Palla (1997) from $z=$1100 to $z$=20, and the asymptotic, constant value $x_{rel}(z)$=3$\times{}10^{-4}$ for $z<20$. We found that the differences obtained considering other estimates of $x_{rel}(z)$ -- as those derived in Jones \& Wyse (1985), Seager, Sasselov \& Scott (2000), or Shull (2004) -- are negligible.
%%%%%%\begin{itemize}
%%%%%%\item if $1000\geq{}z\geq{}800$, we use the approximation given by Jones \& Wyse (1985) updating it to WMAP cosmological parameters:
%%%%%%\qa\label{eq:jones}
%%%%%%x_{rel}(z)=\frac{206.1\,{}(1+1.45\,{}z/z_{eq})^{1/2}}{z^2}\frac{(\Omega{}_Th^2)^{1/2}}{\Omega{}_bh^2}\nonumber{}\\
%%%%%%\times{}\left[1+2.26\times{}10^4\,{}z\,{}e^{-14.620/z}\right],
%%%%%%\nqa 
%%%%%%where $z_{eq}\simeq{}4\times{}10^4h^2$ is the equivalence redshift, $\Omega{}_T=\Omega{}_{0M}+\Omega{}_\Lambda{}=1$, and $h=0.72$.
%%%%%%\item If $800\geq{}z\geq{}10$, we adopt the asymptotic value $x_{rel}(z)=5\times{}10^{-4}$ (Tegmark, Silk, Rees Blanchard, Abel \& Palla 1997; Seager, Sasselov \& Scott 2000; Shull 2004).
%%%%%%\end{itemize}
The ionized fractions we calculated in this way are shown in Fig.~8. The ionization fraction due to radiatively decaying neutrinos of masses $m_{\nu{}s}\lesssim{}14$ keV (higher masses are prohibited by X-ray measurements) is always $x<0.1$. For example, the ionization fraction produced by a sterile neutrino of 2 keV (14 keV) exceeds the relic ionization fraction at redshift $z\sim{}50$ ($z\sim{}200$) and increases up to $x\sim{}0.003$ ($x\sim{}0.01$) at redshift $z=10$.
%As we can see the contribution of sterile neutrinos is higher than that due to residual ionization for redshift $z<450$, and becomes significant ($x\gtrsim{}0.01$) at redshift $z\sim{}30$; but it never approaches the unity, at least for the neutrino masses considered (those permitted by power spectrum distortion and X-ray background).
%%%%%%%%%%%%%%%%%%%%%%%%%%%%%%%%%%% FIGURE 8 %%%%%%%%%%%%%%%%%%%%%%%%%%%%%%%%%%
\begin{figure}
\center{{
\epsfig{figure=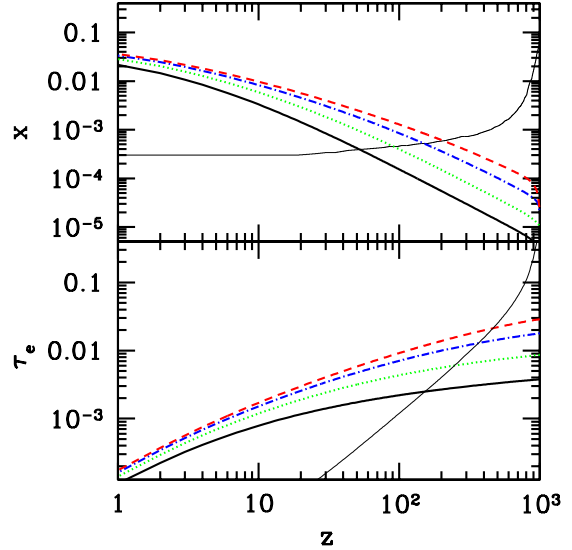,height=8cm}
}}
\caption{\label{fig:fig8} 
Ionization fraction ({\it top panel}) and Thomson optical depth ({\it bottom panel}) due to radiatively decaying neutrinos of masses 2 ({\it solid line}), 4 ({\it dotted line}), 8 ({\it dot-dashed line}) and 14 keV ({\it dashed line}). The {\it solid thin line} indicates: in the {\it top panel} the relic ionization fraction (Tegmark et al. 1997); in the {\it bottom panel} the Thomson optical depth due to  relics.
} 
\end{figure}
%%%%%%%%%%%%%%%%%%%%%%%%%%%%%%%%%%%%%%%%%%%%%%%%%%%%%%%%%%%%%%%%%%%%%%%%%%%%%%%

It is then possible to derive the Thomson optical depth $\tau{}_e$ due to radiatively decaying neutrinos using the 
standard expression 
\q \label{eq:eq28}
\tau{}_e=\int_0^z{\ud{}}z\,{}\frac{{\ud{}}t}{{\ud{}}z}\,{}c\,{}\sigma{}_T\,{}n_{e}(z),
\nq
where $\sigma{}_T$ is the Thomson cross section, ${\ud{}}t/{\ud{}}z=H_0^{-1}[\Omega{}_{0M}(1+z)^3+\Omega{}_\Lambda{}]^{-1/2}$ and $n_{e}(z)=n_{H+}(z)\equiv{}x(z)\,{}n_{H}(0)\,{}(1+z)^3$. The results are shown in Fig.~8.
%%%%%% If as the upper limit of integration we chose $z=136$, that is the redshift where the temperature of relic free electrons deviates from the temperature of the CMB (Barkana \& Loeb 2001), because  we want that the value of $\tau{}_e$ we are calculating does not depend on the relics, we find the results reported in Table 1.
For an integration upper limit $z=1000$ (i.e. corresponding to the low redshift boundary of the last scattering surface), we find the results reported in Table 1.  The contribution of radiatively decaying neutrinos to ionization fraction is 
negligible, being in the range $3.77\times{}10^{-3} \le \tau{}_e \le 2.86\times{}10^{-2}$ for $m_{\nu{}s}=2-14$ keV, 
i.e.  much lower than the WMAP best value ($\tau{}_e=0.17\pm{}0.04$). 
% As a comparison we have to notice that the $\tau{}_e$  due to relic electrons is $\tau{}_e=0.1081$, much higher than the contribution of sterile neutrinos.}
%%%%%%%%%%%%%%%%%%%%%%%%%%%%%%%%%%% TABLE 1 %%%%%%%%%%%%%%%%%%%%%%%%%%%%%%%%%%%
\begin{table}
\begin{center}
\caption{Thomson optical depth due to radiatively decaying and pion-decaying sterile neutrinos (integrated from $z$=1000 to $z$=0).
}
\begin{tabular}{cc}
\hline
\hline
\vspace{0.1cm}
$m_{\nu{}s}$ (keV) & $\tau{}_e$\\
%\hline
2           & 3.77$\times{}10^{-3}$ \\ 
4           & 8.61$\times{}10^{-3}$ \\ 
6           & 1.33$\times{}10^{-2}$ \\ 
8           & 1.78$\times{}10^{-2}$ \\ 
10          & 2.18$\times{}10^{-2}$ \\ 
12          & 2.54$\times{}10^{-2}$ \\ 
14          & 2.86$\times{}10^{-2}$ \\ 
\hline 
\vspace{0.1cm}
$m_{\nu{}s}$ (MeV) & $\tau{}_e$\\
%\hline
150         & 5.81$\times{}10^{-3}$ \\ 
215         & 2.14$\times{}10^{-3}$ \\ 
250         & 1.71$\times{}10^{-3}$ \\ 
300         & 1.35$\times{}10^{-3}$ \\ 
350         & 1.13$\times{}10^{-3}$ \\ 
400         & 9.85$\times{}10^{-4}$ \\ 
450         & 8.78$\times{}10^{-4}$ \\ 
500         & 7.96$\times{}10^{-4}$ \\ 
\hline
\end{tabular}
\end{center}
%\begin{flushleft}
%{\footnotesize $^{a}$Totani \& Takeuchi model rescaled to the Spitzer data.}\\
%{\footnotesize $^{b}$(K) and (W) indicate the ZL subtraction obtained using Kelsall's model and Wright's model, respectively.}\\
%\end{flushleft}
\label{tab_1}
\end{table}
%%%%%%%%%%%%%%%%%%%%%%%%%%%%%%%%%%%%%%%%%%%%%%%%%%%%%%%%%%%%%%%%%%%%%%%%%%%%%%%

\subsection{Pion decay channel}
To calculate the ionization fraction due to CMB photons Compton-scattered by electrons produced by heavy sterile neutrinos we use again eq. (\ref{eq:eq18}). In this case:
\qa \label{eq:eq29}
\frac{\ud{}N}{{\ud{}}E\,{}\ud{}A\,{}{\ud{}}{\it t}}=\frac{1}{4\pi{}}\frac{8\pi{}^2r_0^2}{h_{Pl}^3c^2}\,{}F(p)\,{}(k_BT_0)^{(p+5)/2}E_{obs}^{-(p+1)/2}\nonumber{}\\
\times{}\,{}\frac{4.8}{\tau{}\,{}\left[\left(\frac{m_{\nu{}s}}{m_{\pi{}}}\right)^2-1\right]}\,{}\frac{3\,{}n_b}{11\,{}\eta{}}\frac{1}{\int^{\gamma{}_{max}}_{\gamma{}_{min}}\gamma{}^{-p}\ud{}\gamma{}}\,{}\frac{c}{H_0}\nonumber{}\\
%\times{}\,{}\frac{4.8\times{}10^{-15}}{\tau{}\,{}\left[\left(\frac{m_{\nu{}s}}{m_{\pi{}}}\right)^2-1\right]}\,{}\frac{3\,{}n_b}{11\,{}\eta{}}\frac{1}{\int^{\gamma{}_{max}}_{\gamma{}_{min}}\gamma{}^{-p}\ud{}\gamma{}}\,{}\frac{c}{H_0}\nonumber{}\\
\times{}\int^{z_{decay}}_{z} \frac{(1+z)\,{}e^{-t(z)}\,{}\left(e^{t_C(z)/\tau{}}-1\right)}{\left[\Omega{}_{0M}(1+z)^3+\Omega{}_{\Lambda{}}\right]^{1/2}}\,{}e^{-\tau_{abs}(z,z_{decay})}\,{}{\ud{}}z
\nqa
where $\tau_{abs}(z,z_{decay})$ is given by the eq. (\ref{eq:eqtau}). Following the same procedure as in previous Section (see also Appendix~A), we find:
\qa \label{eq:eq30}
\frac{x^2(z)}{(1-x(z))}=\frac{4\pi{}}{\alpha{}(T)\,{}n_{H}(0)\,{}(1+z)^2}\int^{E_{em}/(1+z)}_{E_{th}/(1+z)}\frac{\ud{}N}{{\ud{}}E\,{}\ud{}A\,{}{\ud{}}{\it t}}\nonumber{}\\
\times{}\,{}\sigma{}_{E(z)}\,{}\left[1+\phi{}(x(z))\frac{E_{obs}(1+z)}{E_{th}}\right]\,{}{\ud{}}E_{obs},
\nqa
where $\ud{}N/{\ud{}}E\,{}\ud{}A\,{}{\ud{}}{\it t}$ comes from the eq. (\ref{eq:eq29}), $\alpha{}(T)$, $\sigma{}_{E(z)}$, $E_{th}$ and $\phi{}(x(z))$ are the same as defined for radiatively decaying neutrinos. We adopted lifetime $\tau{}=10^{15}$s as suggested by HH. Because of the constraints derived from the HXRB, only masses $215\lesssim{}m_{\nu{}s}\lesssim{}500$ MeV are allowed assuming this lifetime.

%Applying the eq. \ref{eq:eq30}, we calculated the ionization fraction due to CMB photons Compton scattered by electrons produced by neutrino decay. 
The results shown in Fig.~9 imply that the ionization fraction due to pion-decaying neutrinos is negligible at high redshift, being always lower or close to the relic one for $z\gtrsim{}30$. At redshifts lower than 20, the contribution of 
sterile neutrinos becomes more important, albeit $x_e \la 0.1$; as a consequence, the Thomson optical depth 
(Fig.~9 and Table 1) is very low ($\tau{}_e<10^{-2}$). Even for $m_{\nu{}s}=150$ MeV with a lifetime $\tau{}=10^{15}$ s (which is however forbidden by HXRB limits), we only obtain $\tau_e=5.8\times{}10^{-3}$ (Table 1). This result discards 
pion-decaying neutrinos as a source of reionization, in agreement with the results of Pierpaoli (2004). 
 
%%%%%%%%%%%%%%%%%%%%%%%%%%%%%%%%%%% FIGURE 9 %%%%%%%%%%%%%%%%%%%%%%%%%%%%%%%%%%
\begin{figure}
\center{{
\epsfig{figure=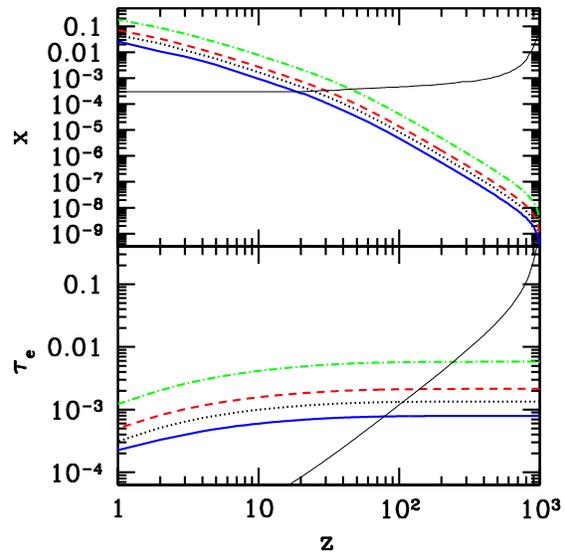,height=8cm}
}}
\caption{\label{fig:fig9} 
Ionization fraction ({\it top panel}) and Thomson optical depth ({\it bottom panel}) due to pion-decaying sterile neutrinos of masses 500 ({\it solid line}), 300 ({\it dotted line}), 215 ({\it dashed line}) and 150 MeV ({\it dot-dashed line}). 
The {\it solid thin line} indicates: in the {\it top panel} the relic ionization fraction (Tegmark et al. 1997); in the {\it bottom panel} the Thomson optical depth due to  relics.
} 
\end{figure}
%%%%%%%%%%%%%%%%%%%%%%%%%%%%%%%%%%%%%%%%%%%%%%%%%%%%%%%%%%%%%%%%%%%%%%%%%%%%%%%
%%%%%%%%%%%%%%%%%%%%%%%%%%%%%%%%%%% TABLE 2 %%%%%%%%%%%%%%%%%%%%%%%%%%%%%%%%%%%
%%%%%%%\begin{table}
%%%%%%%\begin{center}
%%%%%%%\caption{Thomson optical depth due to pion-decaying neutrinos (at $z=136$).
%%%%%%%}
%%%%%%%\begin{tabular}{cc}
%%%%%%%\hline
%%%%%%%\hline
%%%%%%%$m_{\nu{}s}$ (MeV) & $\tau{}_e$\\
%%%%%%%\hline
%%%%%%%150           & 0.00280 \\ 
%%%%%%%250           & 0.00240 \\
%%%%%%%500         & 0.00237 \\
%%%%%%%\hline
%%%%%%%\end{tabular}
%%%%%%%\end{center}
%%%%%%%\label{tab_1}
%%%%%%%\end{table}
%%%%%%%%%%%%%%%%%%%%%%%%%%%%%%%%%%%%%%%%%%%%%%%%%%%%%%%%%%%%%%%%%%%%%%%%%%%%%%%
\subsection{Comparison with previous results}
The values of the ionization fraction and of the Thomson optical depth derived in Section 4.2 partially disagree with the scenario drawn by HH, who suggest an early ionization due to pion-decaying neutrinos. In this Section we clarify the possible sources of such discrepancy. 

Let us consider first the flux of ionizing photons in the two models. This flux can be derived by integrating eq. (\ref{eq:eq29}) from $E_{th}/(1+z)$ to $E_{em}/(1+z)$. At redshift $z=20$ (the reference redshift in HH) we find that the ionizing photons flux is $\sim{}30$ cm$^{-2}$ s$^{-1}$. In the case studied by of HH, the flux of ionizing photons can be derived from their equation (10), i.e. $n_b=\epsilon{}\chi{}n_e^{prod}$, 
where $n_b$ is the number density of (re)ionized hydrogen atoms, $\chi{}=E_e/E_{th}$, $n_e^{prod}$ is the number density of electrons produced by neutrino decays, and $\epsilon{}\sim{}0.3$ is the fraction of the initial energy of the secondary photoelectrons spent in collisional ionizations (Shull \& van Steenberg 1985). In particular, we can write:
\q \label{eq:eq31}
I=c\,{}\epsilon{}\,{}\chi{}n_e^{prod}
\nq
At $z=20$, for $E_e=100$ MeV, $I\sim{}9.7\times{}10^3$ cm$^{-2}$ s$^{-1}$, that is a difference of more than 2 orders of magnitude with respect to our result. However, if we take $\epsilon{}=\chi{}=1$,
%neglect $\epsilon{}$ and $\chi{}$ in equation (\ref{eq:eq31}), 
we obtain $I\sim{}8.0\times{}10^{-3}$, which is more than 3  orders of magnitude lower than our value. 
%Then, the assumption that the primary electron gives $\chi{}$ ionizations, even if corrected by a factor $\epsilon{}$, is crucial. 
Then, the assumption that an electron of energy $E_e$ is able to produce $\sim{}\epsilon{}\,{}E_e/E_{th}$ ionizations is crucial, and, we think, quite optimistic. For example, a CMB photon Compton-scattered by such electron at $z\gtrsim{}20$ will have an energy much higher than $E_{th}$, and, because $\sigma{}_E\propto{}(E/E_{th})^{-3}$, it is unlikely that this photon  immediately interacts with hydrogen atoms. In addition, this high energy photon can be involved in many other processes, like pair production, and its energy does not entirely contribute to reionization.

However, even under the assumption that  a primary electron yields $\chi{}$ ionizations, additional physics must be considered. 
HH ({\it i}) do not consider the effect of recombinations, ({\it ii}) assume that the density of sterile neutrinos is constant, ({\it iii}) neglect the 
gas clumping factor, ({\it iv}) 
%neglect the fact that only a fraction of the total radiation is immediately absorbed by hydrogen atoms.
do not weight the photoionization rate by the cross section $\sigma{}_{E(z)}$.
 We therefore add these processes to HH's calculation.\\
The number of recombinations ($N_{rec}$) and of ionizations ($N_{ion}$) per unit volume can be written as:
\q \label{eq:eq32}
\begin{array}{l}
N_{rec}=\alpha{}\,{}C\,{}x^2n_b^2(1+z)^6\\
N_{ion}=(1-x)\,{}\frac{{\ud}n_{e}}{{\ud}t}\,{}f_{abs}\,{}\epsilon{}\,{}\chi{}
\end{array}
\nq
where $\alpha{}=4.18\times{}10^{-13}\,{}\textrm{cm}^{3}\textrm{s}^{-1}$ is the recombination coefficient, $C\equiv{}\langle{}n_H^2\rangle{}/\langle{}n_H\rangle{}^2$($\gtrsim{}1$) is the clumping factor, $x$ the ionization fraction and $n_b$ the present baryon density. $\chi{}$ and $\epsilon{}$ are the same as previously defined. $f_{abs}$ is the fraction of the total radiation which is absorbed at each redshift (to take into account the photoionization cross section).
%, that is  $f_{abs}$ takes into account that only a part of the Compton scattered radiation interacts with hydrogen atoms. 
Finally, ${\ud}n_{e}/{\ud}t$ is the production rate of primary electrons, i.e.
${\ud}n_{e}/{\ud}t=n_e\,{}\tau{}^{-1}(1+z)^3\,{}e^{-t(z)/\tau{}}$,
where $n_e=4.8\,{}n_a\,{}\tau{}^{-1}\,{}((m_{\nu{}s}/m_{\pi{}})^2-1)^{-1}$ and $n_a=3/11\,{}n_b\,{}\eta{}^{-1}$ cm$^{-3}$ (see Sec. 2.2).\\
Then, imposing $N_{rec}=N_{ion}$ we find:
\q \label{eq:eq33}
%\frac{x^2}{(1-x)}=\frac{1.3}{\alpha{}\,{}C\,{}n_b\,{}\eta{}\,{}(1+z)^3}\frac{e^{-t(z)/\tau{}}}{\tau{}^2\,{}((m_{\nu{}s}/m_{\pi{}})^2-1)}\,{}f_{abs}\,{}\epsilon{}\,{}\chi{}
\frac{x^2}{(1-x)}=\frac{1.31\,{}e^{-t(z)/\tau{}}\,{}f_{abs}\,{}\epsilon{}\,{}\chi{}}{\alpha{}\,{}C\,{}n_b\,{}\eta{}\,{}(1+z)^3\,{}\tau{}^2\,{}((m_{\nu{}s}/m_{\pi{}})^2-1)},
\nq
where $t(z)$ is given by eq. (\ref{eq:eq2}). For comparison with HH, we adopt $\tau{}=10^{15}$ s and $z=20$. At this redshift, we can neglect\footnote{This results in a difference of the order of $10^{-4}$} $(1-x)$, because $x\ll{}1$. As $E_e=1/2\,{}(m_{\nu{}s}-m_\pi{})$ and choosing $m_{\nu{}s}=$215 MeV (i.e. the minimum mass allowed by the HXRB for $\tau{}=10^{15}$ s, corresponding to the maximum $x$), we obtain:
\q \label{eq:eq34}
x\sim{}0.070\,{}\left(\frac{\epsilon{}}{0.3}\right)^{1/2}\,{}f_{abs}^{1/2}\,{}C^{-1/2}
\nq
%We chose the case $m_{\nu{}s}=$215 MeV because this is the minimum mass allowed by the HXRB for a lifetime $10^{15}$ s, and, then, gives the maximum $x$.
Even if $f_{abs}=1$ and $C=1$, an ionization fraction $x(z=20)\sim{}0.07 \ll 1$ is found, i.e. incomplete reionization.   
More realistically, $C$ will be higher than 1, whereas $f_{abs} < 1$. For example, by calculating the difference between the ionizing flux due to Compton scattered CMB photons (see Sec. 4.2) with and without the term $e^{-\tau{}_{abs}}$ (see eq. (\ref{eq:eqtau})), we find $f_{abs}\sim{}0.1$ in the range $500\geq{}z\geq{}10$. Then, we have $x(m_{\nu{}s}=215\textrm{ MeV})\sim{}0.022\left(\frac{\epsilon{}}{0.3}\right)^{1/2}\,{}\left(\frac{f_{abs}}{0.1}\right)^{1/2}\,{}C^{-1/2}$. This value is in good agreement with the results of Pierpaoli (2003), but it is a factor 10 higher than that reported in Fig. 9. This difference can be explained with the fact that  equation (\ref{eq:eq33}) still contains some simplifying assumptions, as the term $\chi{}$ (see previous discussion), instead of a more realistic spectrum for the ionizing radiation.

%The ionizing radiation due to Compton scattered photon is:
%\q \label{eq:eq34}
%I(z)=\int^{E_{em}/(1+z)}_{E_{th}/(1+z)}\frac{\ud{}N}{{\ud{}}E\,{}\ud{}A\,{}{\ud{}}{\it t}}\,{}\sigma{}_{E(z)}\,{}\left[1+\phi{}(x(z))\frac{E_{obs}(1+z)}{E_{th}}\right]\,{}{\ud{}}E_{obs},
%\nq
%where $\frac{\ud{}N}{{\ud{}}E\,{}\ud{}A\,{}{\ud{}}}$ is given in eq. (\ref{eq:eq29}). Integrating eq. (\ref{eq:eq34}) with and without the term $e^{-\tau{}_{abs}}$ (see eq. (\ref{eq:eqtau})), we 
 
\section{Summary}     
We have studied the possible contribution of sterile neutrinos decays to the extragalactic X-ray and infrared-optical 
background, deriving limits on their masses. Such constraints have been then used to assess the impact of ionizing
radiation produced by the decays on cosmic reionization. Both radiatively decaying neutrinos and heavy neutrinos 
decaying into pions and electrons (HH) have been considered.

By calculating the contribution of sterile neutrinos to X-ray background, we have put a strong constraint on the 
neutrino mass. In particular, by requiring that the estimated flux due to sterile neutrinos does not exceed the SXRB 
(Moretti et al. 2003; Dijkstra et al. 2004; Bauer et al. 2004), we found that radiatively decaying neutrinos must have masses 
$m_{\nu{}s}\la$950 keV.
%, whereas for pion-decaying sterile neutrinos $m_{\nu{}s}\ga 190$~MeV. 
Much more stringent limits 
can be found from the comparison with the HXRB. In fact, by requiring that the flux produced by sterile 
neutrino decays does not exceed the HXRB (Bauer et al. 2004), we find that the mass of radiatively decaying neutrinos 
must be  $ m_{\nu{}s}\la\,{}14$~keV.
This constraint on  radiatively decaying neutrinos is very important if combined with the results by Viel et al. (2005),
who found a lower limit $m_{\nu{}s}\gtrsim{}2$ keV from the study of power spectrum. 
%%%%As a caveat, we point out that 
%%%%these estimates might require a correction for the clustering of dark matter.
%, and for pion-decaying  $m_{\nu{}s}\ga$200~MeV. A different X-ray background model in the 
%3-60 keV range (Gruber 1992), provides even stringent constraints, i.e. $m_{\nu{}s}\la 3$~keV for radiatively-decaying, 
%and $m_{\nu{}s}\ga$200~MeV for pion-decaying sterile neutrinos, respectively. 
Pion-decaying neutrino masses in the range  150~MeV $\lesssim{}m_{\nu{}s}\lesssim$500~MeV are allowed if the lifetime is longer than $4\times{}10^{17}$ s. For shorter lifetimes, the minimum mass increases: for example, if $\tau{}=10^{15}$ s, $215\lesssim{}m_{\nu{}s}\lesssim{}500$ MeV.

We have calculated the sterile neutrino decay contribution to the optical and infrared background light. We find  that 
both radiatively and pion-decaying neutrinos give a flux in the optical and NIR range which is several orders of 
magnitude lower than the observed one. 
%As a result,
Then alternative sources, as very massive Pop~III stars (Salvaterra \& Ferrara 2003), must be invoked to explain the detected NIR background excess.
%, as very massive Pop~III stars (Salvaterra \& Ferrara 2003). 

Decaying sterile neutrinos might also be a potential source of cosmological reionization. We derived the ionization 
fraction and the Thomson optical depth produced by radiatively decaying sterile neutrinos and CMB photons 
Compton-scattered by electrons produced by heavy (pion-decaying) neutrinos. Radiatively decaying neutrinos 
produce an optical depth  $\tau_e =(0.4-3)\times{}10^{-2}$, whereas 
the contribution to reionization of pion-decaying neutrinos is even smaller, 
yielding an optical depth $\tau_e \approx 10^{-3}$. 
In conclusion, our calculations suggest that both radiatively- and pion-decaying neutrinos 
are not viable reionization sources.  

\section*{Acknowledgements}
We thank S.~Hansen, A.~Moretti, P.~Tozzi, E.~Ripamonti, P.~Ullio, T.~Schwetz and E.~Pierpaoli for useful discussions.

%Formato per le figure
%\begin{figure}
%\center{{
%\epsfig{figure=name.ps,height=8cm}
%}}
%\caption{\label{name} caption}
%\end{figure}

\onecolumn
\appendix

\section{Ionization fraction by radiatively decaying sterile neutrinos}
%\subsection{Radiatively decaying sterile neutrinos}
The hydrogen ionization fraction, $x$, due to photons emitted by neutrino decays can be derived, under the hypothesis of ionization equilibrium (Osterbrock 1988), as:
\q\label{eq:ap1}
(1-x)\,{}n_H\int_{E_{th}}^{\infty{}}4\pi{}\frac{\ud{}N}{\ud{}{\it E}\,{}\ud{}A\,{}\ud{}{\it t}}\sigma{}_E\,{}{\ud{}}{\it E}=x^2\,{}n_H^2\alpha{}(T),
\nq
where $n_H=n_{H0}+n_{H+}$ is the total hydrogen number density ($n_{H0}$ and $n_{H+}$ being the neutral and ionized hydrogen number density respectively);
%%% considering both neutral ($n_{H0}$) and ionized atoms ($n_{H+}$); 
$x\equiv{}n_{H+}/n_{H}$ is the ionization fraction; $E_{th}=13.6$ eV, $\ud{}N/\ud{}{\it E}\,{}\ud{}A\,{}\ud{}{\it t}$ is the photon flux (units of cm$^{-2}$ s$^{-1}$ sr$^{-1}$ erg$^{-1}$), $\alpha{}(T)=4.18\times{}10^{-13}(T/10^4\textrm{K})^{-0.726}$ cm$^3$ s$^{-1}$ is the recombination coefficient; 
%%%(we assumed $\alpha{}(T=10000\,{}\textrm{K})=4.18\times{}10^{-13}$ cm$^3$ s$^{-1}$, Osterbrock 1988). 
$\sigma{}_E$ is the photo-ionization cross section of hydrogen atoms. 
%We used the following approximation (Osterbrock 1988):
%\q\label{eq:eq19}
%\sigma{}_E=\sigma{}_{th}\left[\beta{}\left(\frac{E}{E_{th}}\right)^{-s}+(1-\beta{})\left(\frac{E}{E_{th}}\right)^{-(s+1)}\right],
%\nq
%where $\sigma{}_{th}=6.30e-18$ cm$^{2}$, $\beta{}$=1.34 and $s$=2.99.

We have derived $\ud{}N/\ud{}{\it E}\,{}\ud{}A\,{}\ud{}{\it t}$ from eq. (\ref{eq:eq1}). The specific flux will be:
\q\label{eq:eq19}
\frac{\ud{}N}{\ud{}{\it E}\,{}\ud{}A\,{}\ud{}{\it t}}=\frac{1}{4\pi{}}\frac{c}{H(z)}\frac{n_s(z)}{\tau{}}\int_0^{z'} e^{-t(z')/\tau{}}\,{} e^{-\tau{}_{abs}(z')}
\,{}\frac{\delta{}((1+z')E(z)-E_0)}{(1+z')\,{}\left[(1+z')^3\Omega{}_{M}(z)+\Omega_{\Lambda{}}\right]^{1/2}}\ud{}{\it z'},
\nq
%%%where $z'$ and $0$ are redshifts referred to the frame of the hydrogen atom and correspond, respectively, to $z_{em}$ (the redshift at which the neutrino decays, emitting a photon) and to $z$ in the frame of an observer at redshift 0. Then, the relation between these two reference frames can be written as $(1+z')=(1+z_{em})/(1+z)$. As in Section 2, $E_0$ is the energy of the photon when it was emitted.\\
where, in the reference frame of the hydrogen atom, $z'$ and $0$ are respectively the redshift at which the sterile neutrino decays emitting a photon and the redshift at which this photon ionizes the hydrogen atom.
Passing from the reference frame of the hydrogen atom to the reference frame of an observer (at redshift 0), the redshifts $z'$ and $0$  respectively correspond to $z_{em}$ and $z$, through the relation $(1+z')=(1+z_{em})/(1+z)$.
 As in Section 2, $E_0$ is the energy of the photon when it was emitted.\\
$\tau{}_{abs}(z')$ takes into account the fact that a photon emitted by a sterile neutrino will be absorbed, due to hydrogen photo-ionization, after a certain mean free path $\lambda{}_H$. It is given by:
%%%$\tau{}_{abs}(z')$ is the optical depth of the photon, due to photo-ionization of hydrogen, a term that, if we want to derive the Thomson optical depth, we can no more neglect. $\tau{}_{abs}(z')$ is given by:   
\q\label{eq:eqtau}
\tau{}_{abs}(z')\equiv{}\int_0^{z'}\lambda_H^{-1}(\tilde{z})\,{}\frac{\ud{}{\it l}}{\ud{}\tilde{{\it z}}}\ud\tilde{{\it z}}
\nq
where $\ud{}{\it l}$ is the proper distance element. The mean free path of the photon, $\lambda{}_H$, can be expressed as:
\qs
\lambda{}_H(z)\equiv{}\frac{1}{n_{H0}(z)\,{}\sigma{}_{E(z)}}=\frac{1}{n_{H0}(0)\,{}(1+z)^{3}(1-x(z))\,{}\sigma{}_{E(z)}}.
\nqs
 Because $n_s(z)=n_s\,{}(1+z)^3$ 
%%%(where $n_s$ is the sterile neutrino density today)
 and taking into account that $\delta{}((1+z')E(z)-E_0)\neq{}0$ if and only if $(1+z')=E_0/E(z)$, eq. (\ref{eq:eq19}) becomes
\q\label{eq:eq20}
\frac{\ud{}N}{\ud{}{\it E}\,{}\ud{}A\,{}\ud{}{\it t}}=\frac{1}{4\pi{}}\frac{c}{H(z)}\frac{n_s\,{}(1+z)^3}{\tau{}}\,{} e^{-t(z')/\tau{}}\,{} e^{-\tau{}_{abs}(z')}
\,{}\frac{1}{E_0\,{}\left[\left(\frac{E_0}{E(z)}\right)^3\Omega{}_{M}(z)+\Omega_{\Lambda{}}\right]^{1/2}}
\nq
Substituting $E(z)=E_{obs}(1+z)$ ($E_{obs}$ being the energy of the photon today), $H(z)=H_0[\Omega_{0M}(1+z)^3+\Omega{}_{\Lambda{}}]^{1/2}$ and $\Omega{}_M(z)=\frac{\Omega_{0M}(1+z)^3}{[\Omega_{0M}(1+z)^3+\Omega{}_{\Lambda{}}]}$, we obtain
\q\label{eq:eq21}
\frac{\ud{}N}{\ud{}{\it E}\,{}\ud{}A\,{}\ud{}{\it t}}=\frac{1}{4\pi{}}\frac{c}{H_0[\Omega_{0M}(1+z)^3+\Omega{}_{\Lambda{}}]^{1/2}}
\,{}\frac{n_s\,{}(1+z)^3\,{}e^{-t(z')/\tau{}}\,{} e^{-\tau{}_{abs}(z')}}{\tau{}\,{}E_0\,{}\left[\left(\frac{E_0}{E_{obs}}\right)^3\frac{\Omega{}_{0M}}{[\Omega_{0M}(1+z)^3+\Omega{}_{\Lambda{}}]}+\Omega_{\Lambda{}}\right]^{1/2}},
\nq
where, assuming that $\Omega_\Lambda{}<<(1+z')^3\Omega{}_M(z)$, $\Omega_\Lambda{}<(1+z)^3\Omega{}_{0M}$ and $\Omega{}_M(z)\sim{}1$, and substituting $(1+z')=(1+z_{em})/(1+z)$ and $(1+z_{em})=(E_0/E_{obs})$, we can find
\q\label{eq:eq22}
t(z')\simeq{}\frac{2}{3}H_0^{-1}\Omega{}_{0M}^{-1/2}(E_0/E_{obs})^{-3/2}=t(E_0/E_{obs})
\nq
Similarly, we can also show that
\q\label{eq:eq22tau}
\tau{}_{abs}(z')=\int_0^{z'}\lambda_H^{-1}(\tilde{z})\,{}\frac{\ud{}{\it l}}{\ud{}\tilde{{\it z}}}\ud\tilde{{\it z}}
%%%%%\nonumber{}\hspace{2.6cm}\\
=\int_z^{z_{em}}\lambda_H^{-1}(\tilde{z})\,{}\frac{\ud{}{\it l}}{\ud{}\tilde{{\it z}}}{\ud}\tilde{z}=\tau{}_{abs}(E_0/E_{obs})
\nq
Finally we can write
\q\label{eq:eq23}
\int_{E_{th}}^{\infty{}}4\pi{}\frac{\ud{}N}{\ud{}{\it E}\,{}\ud{}A\,{}\ud{}{\it t}}\sigma{}_E\,{}{\ud{}}E=
%%%%%\nonumber{}\hspace{3.5cm}\\
\frac{c}{H_0}\int_{E_{th}}^{E_0}\frac{n_s\,{}(1+z)^3}{\tau{}\,{}[\Omega_{0M}(1+z)^3+\Omega{}_{\Lambda{}}]^{1/2}}
%%%%%\nonumber{}\hspace{2.8cm}\\
\,{}\frac{\sigma{}_{E(z)}\,{}e^{-t(E_0/E_{obs})/\tau{}}\,{}e^{-\tau{}_{abs}(E_0/E_{obs})}}{E_0\,{}\left[\left(\frac{E_0}{E_{obs}}\right)^3\frac{\Omega{}_{0M}}{[\Omega_{0M}(1+z)^3+\Omega{}_{\Lambda{}}]}+\Omega_{\Lambda{}}\right]^{1/2}}\,{}{\ud{}}E(z)\hspace{0.1cm}
%\int_{E_{th}}^{\infty{}}4\pi{}\frac{\ud{}N}{\ud{}E\,{}\ud{}A\,{}\ud{}t}\sigma{}_E\,{}\ud{}E=\nonumber{}\hspace{3.5cm}\\
%\frac{c}{H_0}\int_{E_{th}}^{E_0}\frac{1}{[\Omega_{0M}(1+z)^3+\Omega{}_{\Lambda{}}]^{1/2}}\nonumber{}\hspace{2.7cm}\\
%\times{}\frac{n_s\,{}(1+z)^3}{\tau{}}\,{}\frac{e^{-t(E_0/E_{obs})/\tau{}}\,{}e^{-\tau{}_{abs}(E_0/E_{obs})}}{E_0}\nonumber{}\hspace{0.25cm}\\
%\times{}\frac{\sigma{}_{E(z)}}{\left[\left(\frac{E_0}{E_{obs}}\right)^3\frac{\Omega{}_{0M}}{[\Omega_{0M}(1+z)^3+\Omega{}_{\Lambda{}}]}+\Omega_{\Lambda{}}\right]^{1/2}}\,{}\ud{}E(z)\hspace{0.1cm}
\nq
where the upper limit of integration, $E_0$, takes into account that the photon energy at redshift $z$ cannot be larger than at redshift $z_{em}$, when it was emitted. We can make a change of integration variable ($E(z)=(1+z)E_{obs}$), obtaining:
\q\label{eq:eq24}
\int_{E_{th}}^{\infty{}}4\pi{}\frac{\ud{}N}{\ud{}{\it E}\,{}\ud{}A\,{}\ud{}{\it t}}\sigma{}_E\,{}{\ud{}}E=
%%%%%\nonumber{}\hspace{3.5cm}\\
\frac{c}{H_0}\int_{E_{th}/(1+z)}^{E_0/(1+z)}\frac{n_s\,{}(1+z)^4}{\tau{}\,{}[\Omega_{0M}(1+z)^3+\Omega{}_{\Lambda{}}]^{1/2}}
%%%%%\nonumber{}\hspace{2.8cm}\\
\,{}\frac{\sigma{}_{E(z)}\,{}e^{-t(E_0/E_{obs})/\tau{}}\,{}e^{-\tau{}_{abs}(E_0/E_{obs})}}{E_0\,{}\left[\left(\frac{E_0}{E_{obs}}\right)^3\frac{\Omega{}_{0M}}{[\Omega_{0M}(1+z)^3+\Omega{}_{\Lambda{}}]}+\Omega_{\Lambda{}}\right]^{1/2}}\,{}{\ud{}}E_{obs}\hspace{0.1cm}
\nq
For $\sigma{}_{E(z)}$ we used the following approximation (Osterbrock 1988):
\q\label{eq:eq25}
\sigma{}_{E(z)}=\sigma{}_{th}
%%%%%\nonumber{}\hspace{6.5cm}\\
\,{}\left[\beta{}\left(\frac{E_{obs}(1+z)}{E_{th}}\right)^{-s}+(1-\beta{})\left(\frac{E_{obs}(1+z)}{E_{th}}\right)^{-(s+1)}\right],
\nq
where $\sigma{}_{th}=6.30\times{}10^{-18}$ cm$^{2}$, $\beta{}$=1.34 and $s$=2.99.

Finally, substituting  eq. (\ref{eq:eq24}) into eq. (\ref{eq:ap1}), reordering, and substituting $n_H(z)=n_H(0)(1+z)^3$, we obtain the expression for the ionization fraction:
\q\label{eq:eq26ap}
\frac{x^2(z)}{(1-x(z))}=\frac{n_s\,{}(1+z)}{\tau{}\,{}\alpha{}(T)\,{}n_H(0)}\,{}\frac{c}{E_0\,{}H_0\,{}[\Omega_{0M}(1+z)^3+\Omega{}_{\Lambda{}}]^{1/2}}
%%%%%\nonumber{}\\
\,{}\int_{E_{th}/(1+z)}^{E_0/(1+z)}\frac{\sigma{}_{E(z)}\,{}e^{-t(E_0/E_{obs})/\tau{}}\,{}e^{-\tau{}_{abs}(E_0/E_{obs})}}{\left[\left(\frac{E_0}{E_{obs}}\right)^3\frac{\Omega{}_{0M}}{[\Omega_{0M}(1+z)^3+\Omega{}_{\Lambda{}}]}+\Omega_{\Lambda{}}\right]^{1/2}}\,{}{\ud{}}E_{obs}
\nq

%%%%%In these calculations we neglected the possibility that the ionizing photon produces a secondary electron which is sufficiently energetic to collisionally ionize other hydrogen atoms. This is an important factor, especially if $E(z)\gg{}E_{th}$. To take in account of it, we have to modify the cross section, introducing a factor $[1+\phi{}(x(z))\,{}E(z)/E_{th}]$, where $\phi{}$ is the parametric function which best fit the behavior of collisional ionizations (as a function of the ionized fraction $x$) in the Monte Carlo simulations of Shull \& van Steenberg (1985). In particular, $\phi{}(x)={\mathcal{C}}\,{}(1-x^a)^b$, where for the ionization of hydrogen ${\mathcal{C}}=0.3908$, $a=0.4092$ and $b=1.7592$.
%%%%%Adding this term in the eq. \ref{eq:eq26} we finally obtain:
%%%%%\qa\label{eq:eq27}
%%%%%\frac{x^2(z)}{(1-x(z))}=\frac{n_s\,{}(1+z)}{\tau{}\,{}\alpha{}(T)\,{}n_H(0)}\,{}\frac{c}{E_0\,{}H_0\,{}[\Omega_{0M}(1+z)^3+\Omega{}_{\Lambda{}}]^{1/2}}
%%%%%\nonumber{}\\
%%%%%\times{}\int_{E_{th}/(1+z)}^{E_0/(1+z)}\frac{e^{-t(E_0/E_{obs})/\tau{}}\,{}e^{-\tau{}_{abs}(E_0/E_{obs})}}{\left[\left(\frac{E_0}{E_{obs}}\right)^3\frac{\Omega{}_{0M}}{[\Omega_{0M}(1+z)^3+\Omega{}_{\Lambda{}}]}+\Omega_{\Lambda{}}\right]^{1/2}}\nonumber{}\\
%%%%%\times{}\quad{}\sigma{}_{E(z)}\,{}\left[1+\phi{}(x(z))\frac{E_{obs}(1+z)}{E_{th}}\right]\,{}{\ud{}}E_{obs}
%%%%%\nqa

%\subsection{Sterile neutrinos decaying into pions and electrons}

\end{document}